\documentclass{article}
\usepackage[preprint]{neurips_2021}

\usepackage{natbib} 
\setcitestyle{square,numbers,comma}
\usepackage{amsmath}
\usepackage{times}
\usepackage{color}
\usepackage{multirow}

\usepackage[utf8]{inputenc} 
\usepackage[T1]{fontenc}    
\usepackage{hyperref}       
\usepackage{url}            
\usepackage{booktabs}       
\usepackage{nicefrac}      
\usepackage{microtype}  
\usepackage{lipsum}		
\usepackage{graphicx}
\usepackage{doi}
\usepackage{amsfonts, amsmath, amsthm, amssymb}
\usepackage{algpseudocode} 
\usepackage{algorithm}
\usepackage{adjustbox}

\DeclareMathOperator*{\ev}{\mathbb{E}}
\usepackage{subfig}

\newcommand{\de}{\,\mathrm{d}}
\newcommand{\vrho}{\mathbr{\rho}}

\newcommand{\hyscoreprime}[1][\vtheta]{\nabla_{\vrho'}\log\nu_{\vrho'}}

\newcommand\NS[1]{ \textcolor{blue}{NS: #1}}
\DeclareMathOperator*{\argmax}{arg\,max}

\title{Bayesian brains and the R\'enyi divergence}

\author{%
  Noor Sajid\thanks{Joint first authors; correspondence to.}\\
  WCHN, University College London, UK \\
  \texttt{noor.sajid.18@ucl.ac.uk} \\
  \And
  Francesco Faccio$^{\ast}$\\
 Swiss AI Lab IDSIA, Switzerland. \\
  \texttt{francesco@idsia.ch} \\
  \And
   Lancelot Da Costa \\
   Imperial College London, UK \& \\
   WCHN, University College London, UK \\
   \texttt{l.da-costa@imperial.ac.uk} \\
  \And
  Thomas Parr \\
  WCHN, University College London, UK \\
  \texttt{thomas.parr.12@ucl.ac.uk} \\
  \And
 J\"{u}rgen Schmidhuber \\
  Swiss AI Lab IDSIA, Switzerland. \\
  \texttt{juergen@idsia.ch} \\
  \And
  Karl Friston \\
  WCHN, University College London, UK \\
  \texttt{k.friston@ucl.ac.uk} \\
}

\begin{document}
\maketitle

\begin{abstract}{
Under the Bayesian brain hypothesis, behavioural variations can be attributed to different priors over generative model parameters. This provides a formal explanation for why individuals exhibit inconsistent behavioural preferences when confronted with similar choices. For example, greedy preferences are a consequence of confident (or precise) beliefs over certain outcomes. Here, we offer an alternative account of behavioural variability using R\'enyi divergences, and their associated variational bounds. R\'enyi bounds are analogous to the variational free energy (or evidence lower bound), and can be derived under the same assumptions. Importantly, these bounds provide a formal way to establish behavioural differences through an $\alpha$ parameter, given fixed priors. This rests on changes in $\alpha$ that alter the bound (on a continuous scale), inducing different posterior estimates, and consequent variations in behaviour. Thus, it looks as if individuals have different priors, and have reached different conclusions. More specifically, $\alpha \to 0^{+}$ optimisation leads to mass-covering variational estimates  and increased variability in choice behaviour. Furthermore, $\alpha \to +\infty$ optimisation leads to mass-seeking variational posteriors, and greedy preferences. We exemplify this formulation through simulations of the multi-armed bandit task. We note that these $\alpha$ parameterisations may be especially relevant, i.e., shape preferences, when the true posterior is not in the same family of distributions as the assumed (simpler) approximate density, which may be the case in many real-world scenarios. The ensuing departure from vanilla variational inference provides a potentially useful explanation for differences in behavioural preferences of biological (or artificial) agents – under the assumption that the brain performs variational Bayesian inference.
}
\end{abstract} 

\section{Introduction}\label{sec::introduction}
The notion that the brain is Bayesian---or more appropriately, Laplacian (\cite{stigler1986history})---and performs some form of inference has attracted enormous attention in neuroscience (\cite{doya2007bayesian,knill2004bayesian}). It takes the view that the brain embodies a model about causes of sensation, that allow for predictions about observations (\cite{Schmidhuber:92ncchunker, schmidhuber1995predictive,dayan1995helmholtz,RN4}) and future behaviour (\cite{Schmidhuber:90diff,friston2017process}). Practically, this involves the optimisation of a free energy functional (or evidence lower bound) (\cite{RN9,friston2017process,RN8}), using variational inference (\cite{blei2017,wainwright2008}), to make appropriate predictions. The free energy functional can be derived from the Kullback-Leibler (KL)-divergence (\cite{kullback1951}), which measures the dissimilarity between true and approximate posterior densities. Under this formulation, behavioural variations can be attributed to altered priors over the (hyper-)parameters of a generative model, given the same (variational) free energy functional (\cite{RN14,RN13}). This has been used to simulate variations in choice behaviour (\cite{Storck:95,RN15,RN14,RN16}) and behavioural deficits (\cite{RN19,RN18}). 

Conversely, distinct behavioural profiles could be attributed to differences in the variational objective, given the same priors. In this paper, we consider this alternative account of phenotypic variations in choice behaviour using  R\'enyi divergences (\cite{Renyi1961, Van2014,phan2019thompson,amari2012,amari2010}). These are a general class of divergences, indexed by an $\alpha$ parameter, of which the KL-divergence is a special case. It is perfectly reasonable to diverge from this special case since variational inference does not commit to the KL-divergence (\cite{wainwright2008}) (indeed, previous work has developed divergence-based lower bounds that give tighter bounds e.g., (\cite{RN67}), yet these may be more difficult to optimise despite being better approximations). Broadly speaking, variational inference is the process of approximating a posterior probability through application of variational methods. This means finding the function (here, an approximate posterior), out of a pre-defined family of functions, that extremizes an objective functional. In variational inference, the key is choosing the objective such that the extreme value corresponds to the best approximation. R\'enyi divergences can be used to derive a (generalised) variational inference objective called the R\'enyi-bound (\cite{Yingzhen2017}). The R\'enyi-bound is analogous to the variational free energy functional and provides a formal way to establish phenotypic differences despite consistent priors. This is accomplished by changes, on a continuous scale, that gives rise to different posterior estimates, and consequent behavioural variations (\cite{Minka2005}). Thus, changing the functional form of the bound will make it will look as if individuals have different priors i.e., have reached different conclusions from the same observations due to the distinct optimisation objective.

It is important to determine whether this formulation introduces fundamentally new differences in behaviour that cannot be accounted for by altering priors under a standard variational objective. Conversely, it may be possible to relate changes in prior beliefs to changes in the variational objective. We investigate this for a simple Gaussian system by examining the relationship between different parameterisations of the R\'enyi bound under fixed priors and the variational free energy under different hyper-priors. It turns out that there is no clear correspondence in most cases. This suggests that differences in behaviour caused by changes in the divergence supplement standard accounts of behavioural differences under changes of priors. 

The R\'enyi divergences depend on an $\alpha$ parameter that controls the strength of the bound and induces different posterior estimates. Consequently, the resulting system behaviour may vary, and point towards different priors that could have altered the variational posterior form. For this, we assume that systems (or agents) sample their actions based upon posterior beliefs, and those posterior beliefs depend on the form of the R\'enyi bound $\alpha$ parameter. This furnishes a natural explanation for observed behavioural variation. To make the link to behaviour, we assume actions are selected -- based on variational estimates -- that maximise the Sharpe ratio (\cite{sharpe1994sharpe}) i.e., a variance-adjusted return. We reserve further details for later sections. Intuitively, under the R\'enyi bound, high $\alpha$ values lead to mass-seeking approximate\footnote{We use approximate and variational posterior interchangeably throughout.} posteriors i.e., greedy preferences for a particular outcome. Conversely, $\alpha \to 0^+$ can result in mode-covering approximate posteriors —- resulting in a greater range of actions for which there are plausible outcomes consistent with prior preferences. Hence, variable individual preferences could be attributed to differences in the variational optimisation objective. This contrasts with standard accounts of behavioural differences, where the precision of some fixed priors is used to explain divergent behaviour profiles under the same variational objective. In what follows, we present, and validate, this generalised kind of variational inference that can explain the implicit preferences of biological and artificial agents, under the assumption that the brain performs variational Bayesian inference.

The paper is structured as follows. First, we provide a primer on standard variational inference using the KL-divergence (section~\ref{sec::vi}). Section~\ref{sec::renyibound} introduces R\'enyi divergences and the derivation for the R\'enyi bound using the same assumptions as the standard variational objective. We then consider what (if any) sort of correspondence exists between the R\'enyi bound and the variational free energy functional —- i.e., the evidence lower bound —- under different priors (section~\ref{sec::post}). In section~\ref{sec::mab}, we validate the approach through numerical simulations of the multi-armed bandit (\cite{auer2002finite, lattimore2020bandit}) paradigm with multi-modal observation distribution. Our simulations demonstrate that variational Bayesian agents, optimising a generalised variational bound (i.e., R\'enyi bound) can naturally account for variations in choice behaviour. We conclude with a brief discussion of future directions and the implications of our work for understanding behavioural variations.

\section{Variational Inference}\label{sec::vi}
Variational inference is an inference scheme based on variational calculus (\cite{RN28}). It identifies the posterior distribution as the solution to an optimisation problem, allowing otherwise intractable probability densities to be approximated (\cite{wainwright2008,RN29}). It works by defining a family of approximate densities over the hidden variables of the generative model (\cite{beal2003,blei2017}). From this, we can use gradient descent to find the member of that variational family that minimises a divergence to the true conditional posterior. This variational density then serves as a proxy for the true density. This formulation underwrites practical applications that characterise the brain as performing Bayesian inference including predictive coding (\cite{schmidhuber1995predictive,Millidge2020,Spratling2017,Whittington2017,perrykkad2020fidgeting}), and active inference (\cite{Storck:95,Costa20210,friston2017process,sajid2021,Tschantz2020}).

\subsection{KL-divergence and the standard variational objective}
To derive the standard variational objective -— known as the variational free energy, or evidence lower bound (ELBO) —- we consider a simple system with two random variables. These are $s \in \mathcal{S}$ denoting hidden states of the system (e.g., it rained last night) and $o \in \mathcal{O}$ the observations (e.g., the grass is wet). The joint density over these variables:

\begin{align}
    p(s,o)=p(o|s)p(s)
\end{align}                                                                     
where, $p(s)$ is the prior density over states and $p(o|s)$ is the likelihood, is called the generative model. Then, the inference problem is to compute the posterior -- i.e., the conditional density -- of the states given the outcomes:

\begin{align}
    p(s|o)=\frac{p(o,s)}{p(o)}.
\end{align}

This quantity contains the evidence, $p(o)$, that can be calculated by marginalising out the states from the joint density. However, the evidence is notoriously difficult to compute, which makes the posterior intractable in practical applications. This problem can be finessed with variational inference\footnote{There are other methods to estimate the posterior that include sampling-based, or hybrid approaches e.g., Markov Chain Monte Carlo (MCMC). However, variational inference is considerably faster than sampling, by employing simpler variational posteriors, which lead to a simpler optimisation procedure (\cite{wainwright2008}).}. For this, we introduce a variational density, $q(\cdot)$ that can be easily integrated. The following equations illustrate how through a few simple moves we can derive the quantities of interest. We assume that both $p(s|o)$ and $q(s)$ are non-zero:

\begin{align}\label{eq::elbo}
     \log p(o) &=\log p(o) + \int_{\mathcal{S}} \log \frac{p(s|o)}{p(s|o)}\de s \\
              &= \int_{\mathcal{S}} q(s) \log p(o)\de s + \int_{\mathcal{S}} q(s) \log \frac{p(s|o)}{p(s|o)} \de s
              = \int_{\mathcal{S}} q(s)\log \frac{p(s,o)}{p(s|o)} \de s \\
              &= \int_{\mathcal{S}} q(s) \log \frac{q(s)}{q(s)}\de s + \int_{\mathcal{S}} q(s)\log p(s,o)\de s + \int_{\mathcal{S}} q(s)\log \frac{1}{p(s|o)}\de s \\
              &= \underbrace{\int_{\mathcal{S}} q(s) \log \frac{1}{q(s)}\de s + \int_{\mathcal{S}} q(s)\log p(s,o)\de s}_{\text{ELBO}} + \underbrace{\int_{\mathcal{S}} q(s)\log \frac{q(s)}{p(s|o)}\de s}_{\text{KL Divergence}} 
\end{align}

The first two summands of the last equality are the evidence lower bound (\cite{Welbourne2011}), and the last summand presents the KL-divergence between the approximate and true posterior. If  $q(\cdot)$ and $p(\cdot)$ are of the same exponential family, then their KL divergence can be computed using the formula provided in (\cite{Huzurbazar1955}). Our variational objective of interest is the free energy functional ($\mathcal{F}$) which upper bounds the negative log evidence. Therefore, we rewrite the last equality:

\small
\begin{align}\label{eq::freeenergy}
    -\log p(o)  &= - \Bigg[\int_{\mathcal{S}} q(s) \log \frac{1}{q(s)}\de s + \int_{\mathcal{S}} q(s)\log p(s,o)\de s + \int_{\mathcal{S}} q(s)\log \frac{q(s)}{p(s|o)}\de s\Bigg] \\
     &\leq - \int_{\mathcal{S}} q(s)\log p(s,o) \de s + \int_{\mathcal{S}} q(s) \log q(s)\de s \\
     &= - \mathbb{E}_{q(s)}[\log p(s,o)] -\mathbb{H}(q(s)) \\
     &= \underbrace{D_{KL} \big[q(s)||p(s)\big]}_{\text{complexity}} - \underbrace{\mathbb{E}_{q(s)} \big[\log p(o|s) \big]}_{\text{accuracy}} \\
     &= \mathcal{F}
\end{align} 
\normalsize

The second last line is the commonly presented decomposition of the variational free energy summands: complexity and accuracy (\cite{friston2017process,sajid2021}). The accuracy term represents how well observed data can be predicted, while complexity is a regularisation term. The variational free energy objective favours accurate explanations for sensory observations that are maximally consistent with prior beliefs.

In this setting, illustrations of behavioural variations i.e., differences in variational posterior estimations can result from different priors over the (hyper-)parameters of the generative model (\cite{Storck:95}), e.g., change in precision over the likelihood function (\cite{RN14}). We reserve description of hyper-priors and their impact on belief updating for section~\ref{sec::post}. 

\section{R\'enyi divergences, and their variational bound}\label{sec::renyibound}
We are interested in defining a (general) variational objective that can account for behavioural variations alternate to a change of priors. For this, we can replace the KL divergence by a general divergence objective, i.e., a non-negative function $D[\cdot||\cdot]$ that satisfies  $D[q(s)||p(s|o)] = 0$ if and only if $q(s)=p(s|o)$ for all $s \in \mathcal{S}$\footnote{Technically, this equality holds up to a set of measure zero.}. For our purposes, we focus on R\'enyi divergences, a general class of divergences that includes the KL-divergence (Table~\ref{table:divergences}). This has the advantage of being computationally tractable, and satisfies many additional properties (\cite{amari2012,Renyi1961,Van2014}). R\'enyi-divergences are defined as (\cite{Yingzhen2017,Renyi1961}):

\begin{align}
   D_{\alpha}\big[p(s|o)||q(s) \big] := \frac{1}{\alpha -1} \log \int_{\mathcal{S}}  p(s|o)^{\alpha} q(s)^{1-\alpha} \de s
\end{align}                  		 

where $\alpha \in \mathbb{R}^+ \setminus \{1\}$.
An analogous definition holds for the discrete case, by replacing the densities with probabilities and the integral by a sum (\cite{Renyi1961}). 
This family of divergences can provide different posterior estimates as the minimum of the divergence with respect to $q$ varies smoothly with $\alpha$. These differences are possible only when the true posterior, e.g., some multi-modal distribution, is not in the same family of distributions as the approximate posterior, e.g., a Gaussian distribution. Note that other (non-R\'enyi) divergences in the literature are also parameterized by $\alpha$, which can lead to confusion: the I divergence, Amari’s $\alpha$-divergence and the Tsallis divergence. All of these divergences are equivalent in that their values are related by simple formulas, see appendix~\ref{appendix::divergences}. This allows the results presented in this paper to be generalised to these divergence families using the relationships in appendix~\ref{appendix::divergences}. 

\subsection{R\'enyi bound}
The accompanying variational bound for R\'enyi divergences can be derived using the same procedures as for deriving the evidence lower bound (Eq.~\ref{eq::elbo}). This gives us the R\'enyi bound introduced in (\cite{Yingzhen2017}):

\begin{align}
    p(o)=\frac{p(o,s)}{p(s|o)}\implies
\end{align}

\begin{align}\label{eq::renyibound}
    &p(o)^{1-\alpha}p(s|o)^{1-\alpha} = p(o,s)^{1-\alpha} \\
    &\int_{\mathcal{S}}  q(s)^{\alpha} p(o)^{1-\alpha} p(s|o)^{1-\alpha}\de s = \int_{\mathcal{S}}  q(s)^{\alpha} p(o,s)^{1-\alpha}\de s \\
    &\log \int_{\mathcal{S}}  q(s)^{\alpha} p(o)^{1-\alpha} p(s|o)^{1-\alpha}\de s= \log \int_{\mathcal{S}}  q(s)^{\alpha} p(o,s)^{1-\alpha}\de s \\
    &\log p(o)^{1-\alpha} = \log \int_{\mathcal{S}}  q(s)^{\alpha} p(o,s)^{1-\alpha}\de s -  \log \int_{\mathcal{S}}  q(s)^{\alpha} p(s|o)^{1-\alpha}\de s \\
    &\log p(o) = \underbrace{\frac{1}{1-\alpha} \log \int_{\mathcal{S}}  q(s)^{\alpha} p(o,s)^{1-\alpha}\de s}_{\text{R\'enyi Bound}}  +  \underbrace{\frac{1}{\alpha-1} \log \int_{\mathcal{S}}  q(s)^{\alpha} p(s|o)^{1-\alpha}\de s}_{\text{R\'enyi Divergence}}\\
    &\log p(o) = -D_{\alpha}[q(s)||p(o,s)] + D_{\alpha}[q(s)||p(s|o)]
\end{align}

We assume that $q(s)$ and $p(s|o)$ are non-zero and $\alpha \in \mathbb{R}^+ \setminus \{1\}$. The negative R\'enyi bound can be regarded as being analogous to the variational free energy objective ($\mathcal{F}$) by providing an upper bound to the negative log evidence (Eq.~\ref{eq::freeenergy}):

\begin{align}
    -\log p(o) &= \frac{1}{\alpha-1} \log \int_{\mathcal{S}}  q(s)^{\alpha} p(o,s)^{1-\alpha}\de s -  \frac{1}{\alpha-1} \log \int_{\mathcal{S}}  q(s)^{\alpha} p(s|o)^{1-\alpha}\de s \label{eq:20}\\
     & \leq \frac{1}{\alpha-1} \log \int_{\mathcal{S}}  q(s)^{\alpha} p(o,s)^{1-\alpha}\de s = D_{\alpha}[q(s)||p(o,s)]\label{eq:21}
\end{align}

Similar to the R\'enyi divergence, we expect variations in the estimation of the approximate posterior with $\alpha$ under the R\'enyi bound. Explicitly, when $\alpha < 1$ the variational posterior will aim to cover the entire true posterior--- this is known as exclusivity (or zero-avoiding) property. In contrast, when $\alpha \to + \infty$ the variational posterior will seek to fit the true posterior at its mode--- this is known as inclusivity (or zero-forcing) mode-seeking behaviour (\cite{Yingzhen2017}). 

Hence, the R\'enyi bound should provide a formal account of behavioural differences through changes in the $\alpha$ parameter. That is, we would expect a natural shift in behavioural preferences as we move from small values to large positive $\alpha$ values, given fixed priors. Section~\ref{sec::mab} demonstrates this shift in preferences in a multi-armed bandit setting. 

\begin{table}[!ht]
\footnotesize
\centering
\begin{adjustbox}{width=1\textwidth,center=\textwidth}
\small
\begin{tabular}{c || c c c}
 \hline
 & R\'enyi Divergence  & R\'enyi Bound &  \\ [0.5ex] 
  $\alpha$ & $D_{\alpha}[p(s|o)||q(s)]$ &$-D_{\alpha}[q(s)||p(s,o)]$ & Comment\\ 

 \hline\hline
  $\alpha \to 1$ & $\int_{\mathcal{S}}  p(s|o) \log \frac{p(s|o)}{q(s)} \de s$ & $-D_{KL}[q(s)||p(s)] + \ev_{q(s)} \log p(o|s)$ & Kullback-Leibler (KL) divergence: $D_{KL}[q||p]$ \\
  & & $-\mathbb{H}[p(s,o)] + \ev_{p(s,o)} \log q(s)$ & or $D_{KL}[p||q]$ \\
    \hline
  $\alpha = 0.5$ & $-2 \log (1-Hel^2(p(s|o), q(s)))$ & $2 \log (\- Hel^2(p(s,o), q(s)))$ & Function of the Hellinger distance or   \\
  & $-2 \log \sqrt{p(s|o)q(s)} \de s$ &$2 \log \sqrt{p(s,o)q(s)} \de s$ & the Bhattacharyya divergence. \\
   &   &  & Both are symmetric in their arguments \\
  \hline
   $\alpha =2$ & $\log \Big[1+{\chi}^2[p(s|o)||q(s)] \Big]$ & $-\log \Big[1+{\chi}^2[q(s)||p(s,o)] \Big]$ & Proportional to ${\chi}^2$-divergence: \\
  &   &  & ${\chi}^2(p,q)=\int_{\mathcal{S}}  \frac{p^2}{q}\de - 1$ \\
\hline
 $\alpha \to  \infty$ & $\log max_{s \in S} \frac{p(s|o)}{q(s)}$ &  $-\log max_{s \in S} \frac{q(s)}{p(s,o)}$ & Minimum description length
  \\ [1ex] 
\hline
\hline
\end{tabular}
\end{adjustbox}
\caption{Examples of (normalised) R\'enyi divergences (\cite{Yingzhen2017,Minka2005,Van2014}) for different values of $\alpha$, and the accompanying R\'enyi bounds. We omit $\alpha \to 0$ because the limit is not a divergence. These divergences have a non-decreasing order i.e., $Hel^2(p,q) \leq D_{\frac{1}{2}}[p||q] \leq D_1 [p||q] \leq D_2 [p||q] \leq \chi^2 (p,q)$ (\cite{Van2014}). }\label{table:divergences}
\end{table}


\normalsize
\section{Variational bounds, precision, and posteriors}\label{sec::post}
It is important to determine whether this formulation of behaviour introduces fundamentally new differences that cannot be accounted for by altering the priors under a standard variational objective. Thus, we compare the R\'enyi bound and the variational free energy on a simple system to see whether the same kinds of inferences can be produced through the R\'enyi bound (Eq.~\ref{eq::renyibound}) with fixed prior beliefs but altered $\alpha$ value and through the standard variational objective (Eq.~\ref{eq::elbo}) with altered prior beliefs. If this were to be the case, we would be able to re-write the variational free energy under different precision hyper-priors as the R\'enyi bound, where hyper-parameters now play the role of the $\alpha$ parameter. If this correspondence holds true, the two variational bounds (i.e., R\'enyi and variational free energy) would share similar optimisation landscapes (i.e., inflection or extrema), with respect to the posterior under some different priors or $\alpha$ value. 

Variations in these hyper-priors speak to different priors, under which agents can exhibit conservative or greedy choice behaviour. Practically, this may be a result of either ($i$) lending one contribution more precision through weighting the log probability under the standard variational objective, or ($ii$) by altering the priors by taking the log of the probability to the power of $\alpha$. To illustrate this equivalence, we consider following. First, we derive the exact variational free energy for a Gaussian system with gamma priors over the variance. Conversely, to derive the exact R\'enyi bound for a similar system, we assume a simple Gaussian parameterisation. The differences in parameterisations allow us to cast the precision prior as being equivalent to the $\alpha$ parameter i.e., one can either alter the precision prior or the $\alpha$ value to evince behavioural differences.

Though the problem setting is simple, it provides an intuition of what (if any) sort of correspondence exists between the R\'enyi bound and the variational free energy functional using different priors.

\subsection{Variational free energy for a Gaussian-Gamma system}\label{sec:veg}
\begin{figure}[!htbp]
  \centering
  \includegraphics[width=0.4\linewidth]{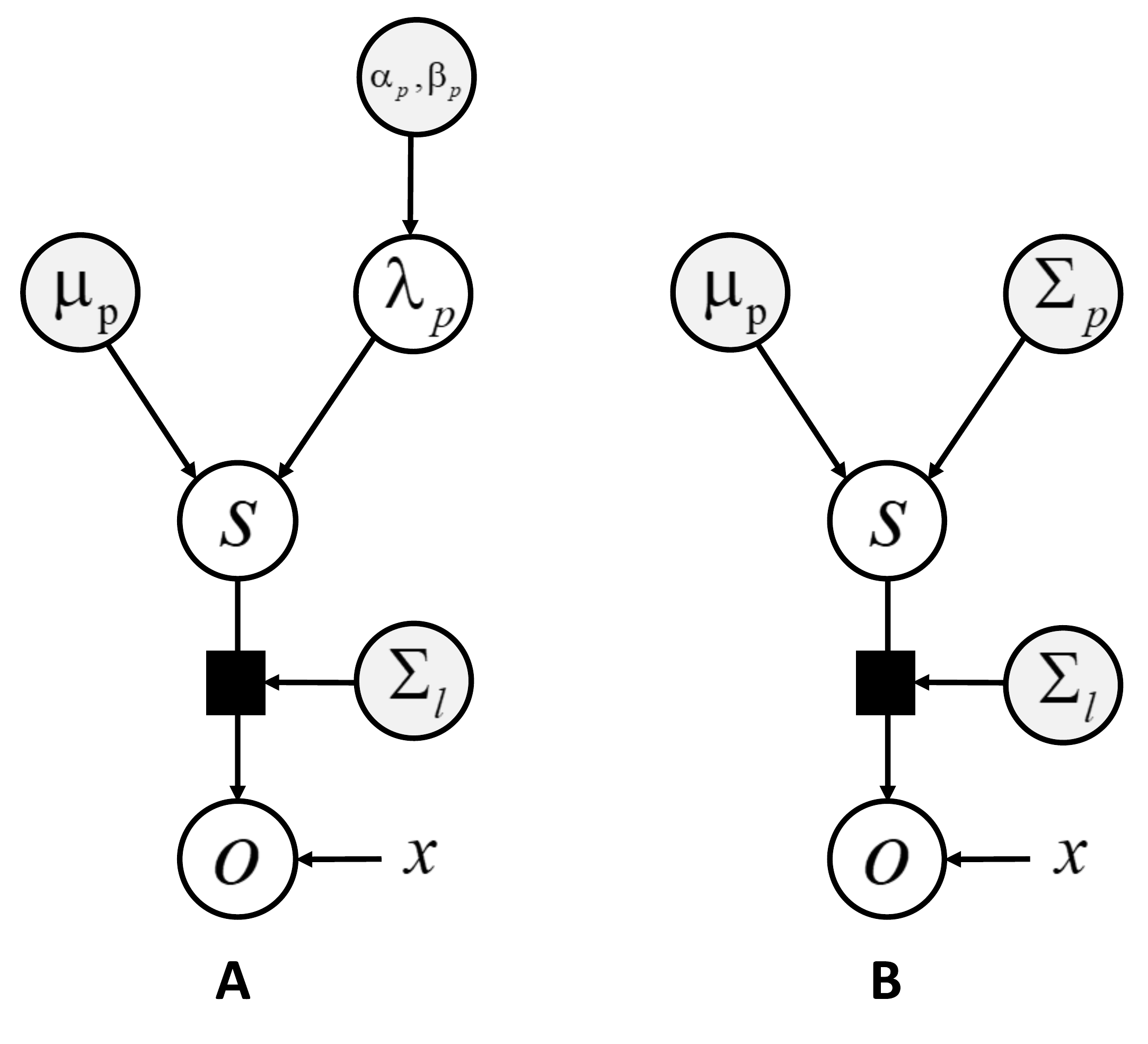}
  \caption{Graphical model for the Gaussian-Gamma (A), and Gaussian (B) system. Here white circles represent random variables, grey circles represent priors and x is the parameter governing the mean. The difference between these models is that in model (A), the precision parameters over hidden states $\lambda_p$ are random variables that follow a Gamma distribution with parameters $\alpha_p, \beta_p$, while in model (B) the precision is held fixed.} 
  \label{fig:architecture} 
\end{figure}

To derive the variational free energy, we consider a simple system with two random variables: $s \in S$ denoting (hidden) states of the system, $o \in O$ the observations (Figure~\ref{fig:architecture} (A)). $\Sigma_k$ is the precision parameter and $x$ the parameter governing the mean. The variational family is parameterised as a Gaussian. This is formalised as: 

\begin{align}
    p(s|\lambda_p) &\sim \mathcal{N}(s;0,(\lambda_p \sigma_p)^{-1})Gam(\lambda_p ;\alpha_p,\beta_p) \\
      p(o|s) & \sim \mathcal{N}(o;sx, \Sigma_l)   \\
    q(s) &\sim \mathcal{N}(s;\mu_q,\Sigma_q)
\end{align}

where $\Sigma_k=(\lambda_k \sigma_k )^{-1}$, $s$ are scalars, $o$ has dimension $n$, and $x$ has dimensionality $n \times 1$. 

We use these quantities to derive the variational free energy (Appendix~\ref{appendix::derivations} for the derivation):

\begin{align}
    -D_{KL}[q(s)||p(s,o)] &= \frac{1}{2}\log\left(\frac{|\Sigma_q|}{(2\pi)^n |\Sigma_p| |\Sigma_l|}\right)\label{eq::kl0} \\
    &- \frac{1}{2}\left(o^T\Sigma_l^{-1}o + \mu_q^2 \Sigma_p^{-1} + \mu_q^2 x^T \Sigma_l^{-1} x -2\mu_q x^T \Sigma_l^{-1} o\right)\label{eq::kl1} \\
    &- \frac{1}{2}\left(\Sigma_q x^T \Sigma_l^{-1} x + \Sigma_q\Sigma_p^{-1} - 1 \right)\label{eq::kl2}\\
    &- \log \frac{\lambda_p^{\alpha_p-1}\beta_p^{\alpha_p}}{\Gamma(\alpha_p)}
    - \lambda_p\beta_p \label{eq::gamma}
\end{align}

Here, Eq.~\ref{eq::gamma} are the additional terms introduced via the Gamma prior. It may seem a little strange for those used to dealing with variational free energy to see it defined in terms of a KL divergence. Often, this notation is reserved for arguments that are both normalised. However, in this paper, we allow the divergences to include un-normalised distributions, facilitating the (unorthodox) expression of variational free energy as a KL divergence.


\subsection{R\'enyi bound for a Gaussian system}
Next, we consider a similar system for deriving the R\'enyi-bound. Unlike for the system in Section~\ref{sec:veg} the densities are parameterised as a Gaussian distribution (Figure~\ref{fig:architecture} (B)):

\begin{align}
    p(s) &\sim \mathcal{N}(s;0,\Sigma_p) \\
      p(o|s) &\sim \mathcal{N}(o;sx,\Sigma_l)  \\
    q(s) &\sim \mathcal{N}(s;\mu_q,\Sigma_q)
\end{align}

where $s$ is a scalar, $o$ has dimension $n$, and $x$ has dimensionality $n \times 1$.
We use these quantities to derive the R\'enyi bound (Appendix~\ref{appendix::derivations} for the derivation):

\begin{align}
    -D_{\alpha}[q(s)||p(s,o)] & = \frac{1}{2}\log\left(\frac{|\Sigma_q|}{(2\pi)^n |\Sigma_p| |\Sigma_l|}\right) \label{eq::rb1}\\
    &- \frac{\alpha}{2(\Sigma_q \Sigma_{\alpha}^{-1})}\left(o^T\Sigma_l^{-1}o + \mu_q^2 \Sigma_p^{-1} + \mu_q^2 x^T \Sigma_l^{-1} x -2\mu_q x^T \Sigma_l^{-1} o\right) \label{eq::rb2}\\
    &- \frac{1}{2(1-\alpha)} \log \left(1+(1-\alpha)(\Sigma_qx^T \Sigma_l^{-1} x + \Sigma_q\Sigma_p^{-1} - 1)    \right)\label{eq::rb3} \\
    &- \frac{1}{2 \Sigma_{\alpha}^{-1}}\left((1-\alpha)\Sigma_p^{-1} o^T \Sigma_l^{-1}o \right) \label{eq::rb4}
\end{align}

where, $\Sigma_{\alpha} := \Big((1-\alpha)(\Sigma_p^{-1} + x^T \Sigma_l^{-1}x) + \alpha \Sigma_q^{-1} \Big)^{-1}$, under the assumption that $\Sigma_{\alpha}$ is positive-definite. Since $\Sigma_{\alpha}$ is a scalar, this is equivalent to satisfying the following condition: $\Sigma_{\alpha} \succ 0 \iff (\alpha -1)(\Sigma_p^{-1} + x^T\Sigma_l^{-1} x)\Sigma_q < \alpha$. Importantly, if $\alpha \leq 1$ the condition is always true for any choice of $\Sigma_q$. However, for $\alpha > 1$ we must impose $\Sigma_q < \frac{\alpha}{\alpha-1} \frac{\Sigma_p}{1 + \Sigma_p x^T \Sigma_l^{-1}x} =  \frac{\alpha}{\alpha-1} \operatorname{Cov}(p(s|o))$ (\cite{burbea1984informative,metelli2018policy}).


\subsection{Correspondence between variational free energy \& the R\'enyi bound}\label{sec::numanal}

Using the derived bounds above, we examine the correspondence between the variational free energy and the R\'enyi bound.

First, we consider the case when $\alpha \to 1$. Here, we expect to find an exact correspondence between the variational free energy and the R\'enyi bound as the R\'enyi divergence tends towards the KL-divergence as $\alpha \to 1$. Our derivations confirm this, upon comparison of the equivalent terms for each objective. The first terms in each objective, Eq.~\ref{eq::kl0} and Eq.~\ref{eq::rb1} are the same. Interestingly, the second term in the R\'enyi bound (Eq.~\ref{eq::rb2}) is a scalar multiple of the second term in variational free energy (Eq.~\ref{eq::kl1}), where the scalar quantity $\frac{\alpha}{\Sigma_q \Sigma_{\alpha}^{-1}}$ tends to $1$ for $\alpha \to 1$. The third term in Eq.~\ref{eq::rb3}, for $\alpha \rightarrow 1$, is a limit of the form $\lim_{x \rightarrow 0} \frac{1}{x}\log(1+xw)=w$, resulting exactly in Eq.~\ref{eq::kl2}. Finally, the last term in the R\'enyi bound tends to zero as $\alpha \to 1$ (Eq.~\ref{eq::rb4}). 

\begin{figure}[!tp]
   \centering
    \subfloat[\centering]{{\includegraphics[width=0.99\textwidth, height=4cm]{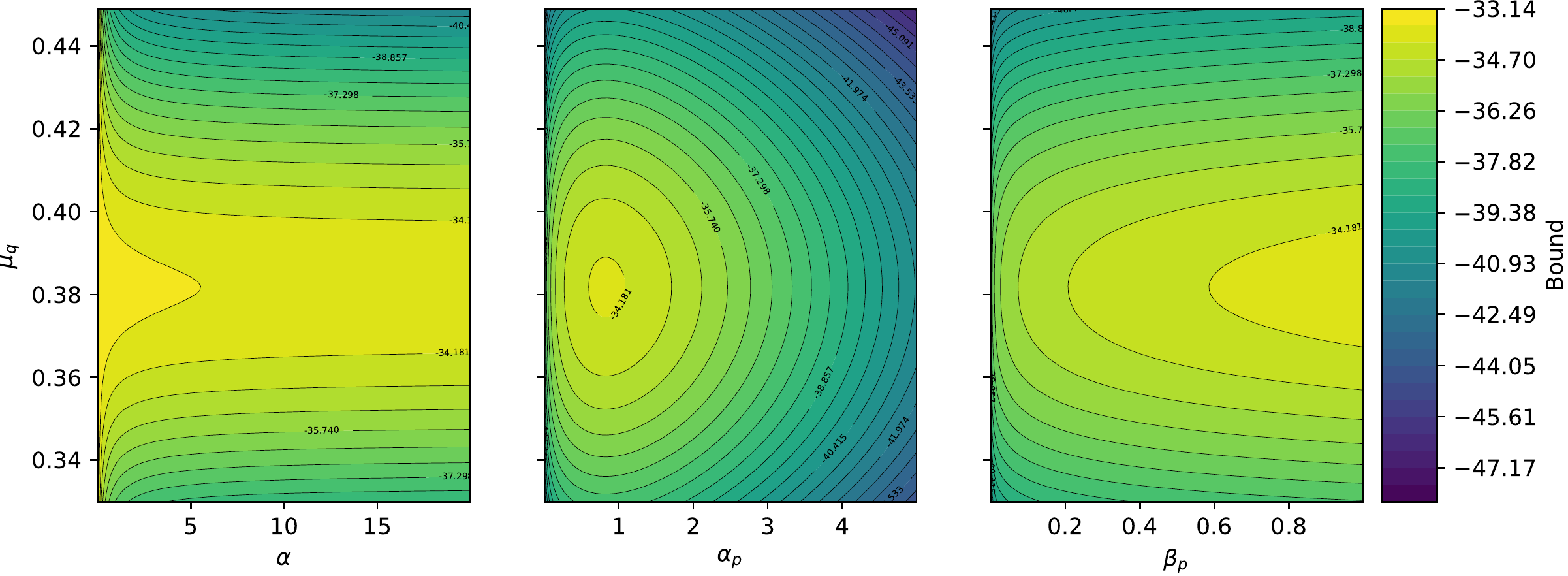} }}%
    \qquad
    \subfloat[\centering]{{\includegraphics[width=0.99\textwidth, height=4cm]{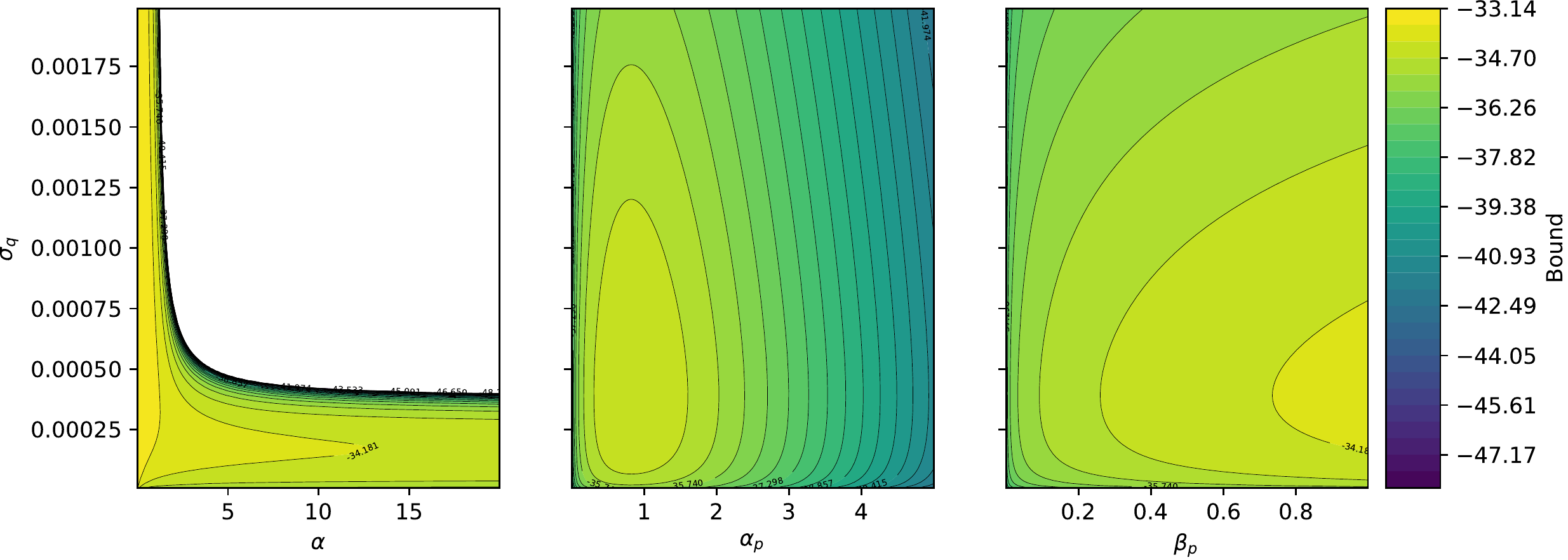} }}%
    \caption{Heat map of variational bounds as a function of estimated sufficient statistics: $\mu_q$ (a) and $\sigma_q$ (b). These graphics plot the optimisation landscape for changing priors or $\alpha$ values. The first column plots the R\'enyi bound, as a function of $\alpha$ on the x-axis and $\mu_q$ (a) or $\sigma_q$ (b) on the y-axis. Similarly, the next two columns plot the free energy, as a function of $\alpha_p$ (center column) or $\beta_p$ (right column) on the x-axis and $\mu_q$ (a) or $\sigma_q$ (b) on the y-axis. The variational bound range from -33 (yellow) to -47 nats (blue). The empty region in (b) for different $\alpha$ values in the Renyi bound is a consequence of the (positive definiteness) constraint imposed on $\Sigma_q$ for $\alpha >1$ restricting the possible values to be $< \frac{\alpha}{\alpha-1} \frac{\Sigma_p}{1 + \sigma_p x^T \sigma_l^{-1}x}$.
    When not varying, hyper-parameters are fixed with $\mu_q=0.4$, $\sigma_q=1e-4$, $\alpha_p=0.8$, $\beta_p=0.8$, $\lambda_p=0.8$, $x=\{r: r=1.1\times n\ , \ \ n\in \{0,1, \dots 19 \} \}$, $y=0.4 \times x$, $\Sigma_l= \mathbb{I}_{20}$. 
    }
    \label{fig:bounds}
\end{figure}

Next, we evaluate the correspondence between the variational free energy and R\'enyi bound when $\alpha \in \mathbb{R}^+ \setminus \{ 1\}$. Now, the $\alpha$ values scale the terms in the R\'enyi bound with Eq.~\ref{eq::rb4} having an influence on the final bound estimate. For comparability, we introduced the Gamma prior to a simple Gaussian system. As shown in Eq.~\ref{eq::gamma}, this introduces additional terms that scale the free energy $\mathcal{F}$. We expect the scaling from the $\alpha$ parameter to have some correspondence to the precision priors in the Gaussian-Gamma system. To assess this, we plot the variational objectives as a function of their estimated sufficient statistics for this simple system (Fig.~\ref{fig:bounds}). The numerical simulation illustrates that optimisation of these objectives, for appropriate priors ($\alpha_p, \beta_p$) or the $\alpha$ value, can lead to (extremely) different variational densities.

Briefly, we do not observe a direct correspondence in the optimisation landscapes (and the variational posterior) for certain priors or $\alpha$ value. These numerical analyses demonstrate that the R\'enyi divergences account for for behavioural differences in a way that is formally distinct from a change in priors, through manipulation of the $\alpha$ parameter. Conversely the standard variational objective could require multiple alterations to the (hyper-)parameters to exhibit a similar functional form in some cases. Further investigation in more complex systems is required to quantify the correspondence (if any) between the two variational objectives. 

\section{Multi-armed bandit simulation}\label{sec::mab}

In this section, we illustrate the differential preferences that arise naturally under the R\'enyi bound. For this, we simulated the multi-armed bandit (MAB)  paradigm (\cite{auer2002finite, lattimore2020bandit}) using $3$ arms. The MAB environment was formulated as a one-state Markov Decision Process (MDP) i.e., the environment remains in the same state independently of agents' actions. At each time-step, the agent could pull one arm and a corresponding outcome (i.e., score) was observed. The agent's objective was to identify, and select, the arm with the highest Sharpe ratio (\cite{sharpe1994sharpe}) through its interactions with the environment across $X$ trials. The Sharpe ratio is a well-known financial measure for risk-adjusted return. It is an appropriate heuristic for action selection because it measures the expected return after adjusting for the variance of the posterior estimate i.e., return to variability ratio. In particular, given the expected return of an arm $R = \ev [R_t]$, the Sharpe ratio is defined as $SR := \frac{\ev[R_t]}{\mathbb{V}[R_t]}$.This heuristic was chosen because it nicely illustrates how changes in  $\alpha$ influence the sufficient statistics of the variational posterior, and ensuing behaviour. 

We modelled each arm with a fixed multi-modal distribution (a mixture of Gaussians) unknown to the agent; characterising this as stationary stochastic bandit setting. Explicitly, this entailed the following parameterisation for each arm: 

\begin{align}\label{eq:mab}
    p(s) &\sim \sum^{2}_{i} \omega_i \mathcal{N}(\mu_i,\Sigma_i) \\
    p(o,s) &\sim \mathcal{N}(s,1.0)p(s) \\
    q(s) &\sim \mathcal{}{N}(\mu_q,\Sigma_q) \\
    \sum^{2}_{i} \omega_i &= 1, \omega_i >0    
\end{align} 

\begin{figure}[!ht]
    \centering
    \includegraphics[width=\textwidth]{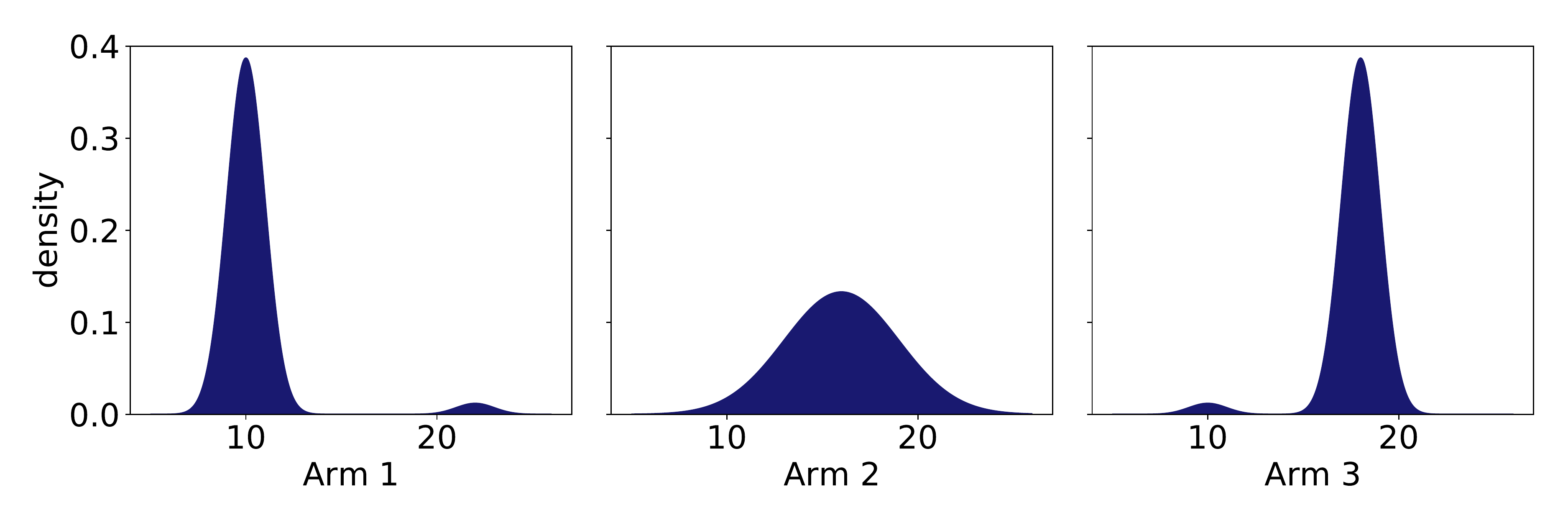}
    \caption{
    Score distribution for each arm. The figures plot the score distributions for each arm. 
    The x-axis is the $s \sim q(s)$ and y-axis the score density. Arm $1$ has a multi-modal distribution of $\mu^1_1 = 10$ ($\Sigma^1_1=1$) and $ \mu^1_2 =22$ ($\Sigma^1_2=1$) with $\omega^1_1=0.97$ and $\omega^1_2=0.03$, respectively. Arm $2$ has a Gaussian distribution with $ \mu^2_1 = 16$ ($\Sigma^2_1 =3$), and Arm $3$ has a multi-modal distribution of $\mu^3_1 = 18$ ($\Sigma^3_1=1$) and $ \mu^3_2 =10$ ($\Sigma^3_2=1$) with $\omega^3_1=0.97$ and $\omega^3_2=0.03$, respectively. }
    \label{fig:mabarm}
\end{figure}

where, $s$ denotes the hidden state over the arm distribution and $o$ the observed return ($R$) from an arm. The variational density $q(s)$ was constrained as a Gaussian with an arbitrary mean and variance, under a mean-field assumption\footnote{That is a fully factorised variational distribution. For further details see (\cite{Minka2005,parr2020modules,sajid2021cancer})}. However, due to the multi-modal prior, the true posterior could take a complex form that might not be in the variational family of distributions. This introduces differences in posteriors that are evident under different R\'enyi bounds. In Fig.~\ref{fig:mabarm}, we show the true distribution for each arm that is unknown to the agent. The Sharpe ratio for arm $1$ was $SR=2.03$; arm $2$ was $SR=1.76$; and arm $3$ was $SR=6.20$. Thus, arm $3$ was the best choice in our paradigm as the arm with the maximal Sharpe ratio. Accordingly, we measured performance using accumulated regret, $\mathcal{R}$, defined as: $\mathcal{R}= \sum^X_{t=1} (SR^*-SR_t)$. Here, $SR^*$ is the maximal Sharpe ratio from arm $3$, and $SR_t$ the Sharpe ratio for the arm pulled at iteration $t$.


\begin{figure}[!htp]
   \centering
    \subfloat[\centering]{{\includegraphics[width=0.45\textwidth, height=4cm]{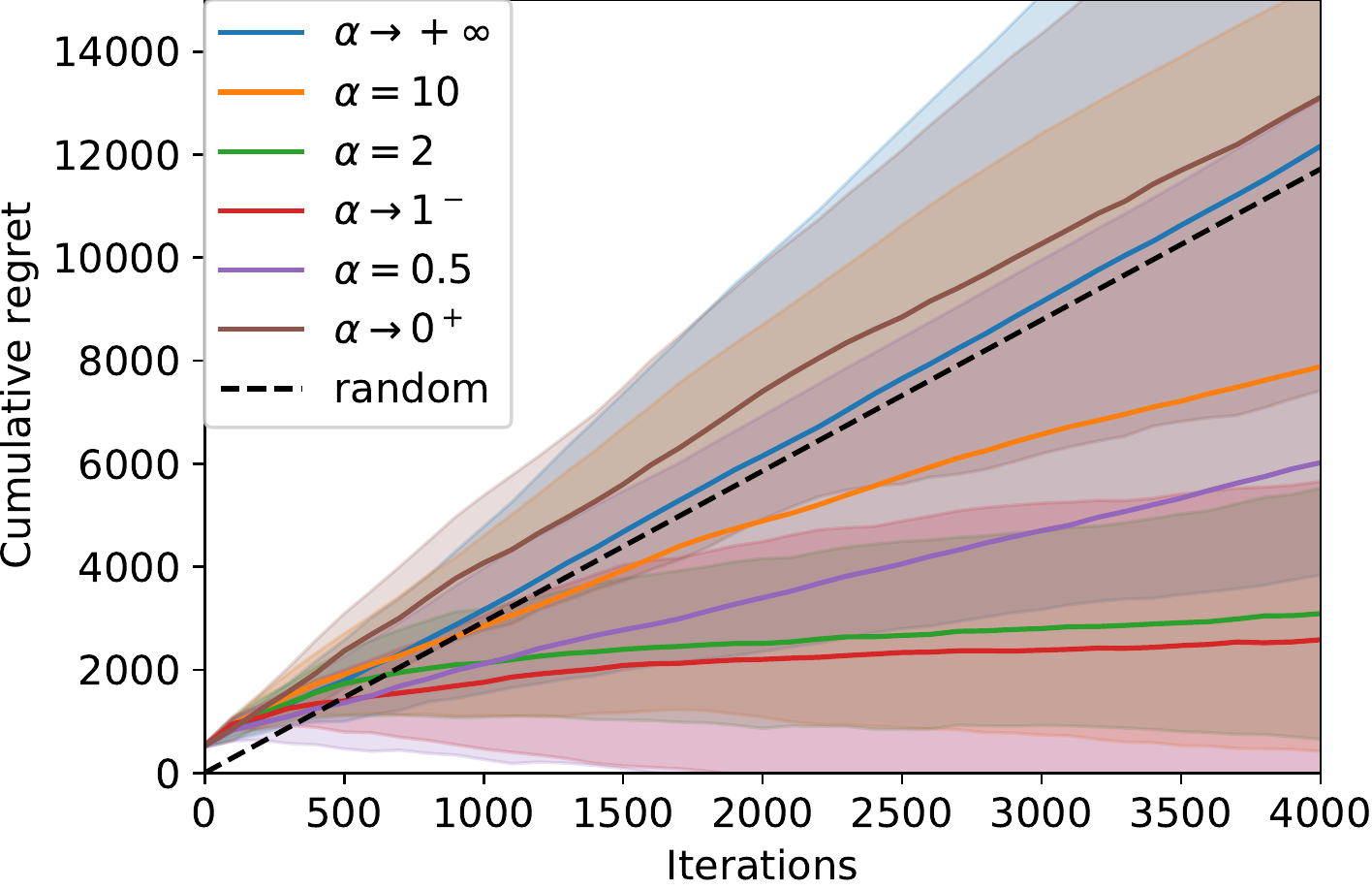} }}%
    \qquad
    \subfloat[\centering]{{\includegraphics[width=0.45\textwidth, height=4cm]{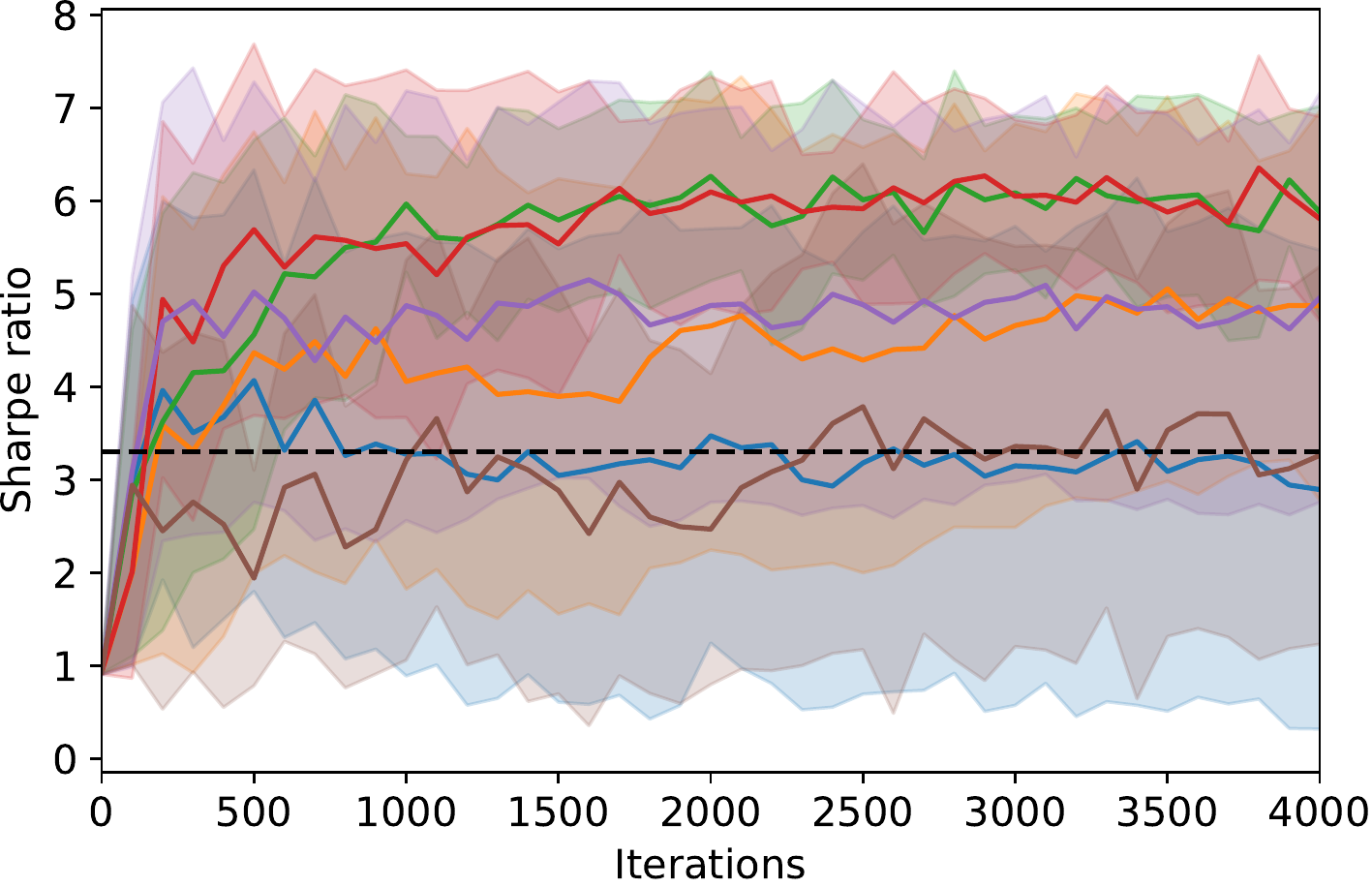} }}%
    \caption{Regret (a) and Sharpe ratio (b) under the R\'enyi bound. (a) The line plot illustrates the cumulative regret across the $4000$ iterations for each agent optimising a particular R\'enyi bound.  The x-axis denotes the iteration and y-axis the accompanying cumulative regret. (b) The line plot illustrates the Sharpe ratio across the $4000$ iterations for each agent optimising a particular R\'enyi bound. The x-axis denotes the iteration and y-axis the Sharpe ratio. Here, blue is for agents optimising R\'enyi Bound for $\alpha \to + \infty$, orange for $\alpha=10$, green for $\alpha=2$, red for $\alpha \to + 1^-$, purple for $\alpha = 0.5$ and brown for $\alpha \to 0^+$. Dashed black line represents regret under a random policy (i.e., any arm). Each agent was simulated 20 times (one standard deviation). In our simulations, the agents with $\alpha \to + 1^-$ and $\alpha=2$ obtained the best performance.}
    \label{fig:mabregret}
\end{figure}

Optimising the R\'enyi bound under different $\alpha$ values led to varying posterior estimates and accompanying behavioural differences manifested by distinct arm choices. To show this, we simulated $6$ agents optimising the R\'enyi bound for distinct $\alpha $ values: $\to + \infty , 10, 2, \to 1^-, 0.5, \to 0^+$ -- across $4000$ iterations, repeated $20$ times for eacha agent. Throughout, the agents selected an arm according to the following strategy. At each iteration, the Sharpe ratio (\cite{sharpe1994sharpe}) was calculated for each arm by dividing a sampled point from the estimated posterior with its variance. The arm with the highest Sharpe ratio was pulled. The agent learnt the score distribution through a memory buffer that stored the previous $1000$ observations. At each iteration the observations in memory were used to optimise the variational posterior estimate.  Appendix~\ref{appendix::code} provides further experimental details. 

The only variable varying across simulations was the $\alpha$ parameter. To assess the performance of each $\alpha$ we plot the accumulated regret, and the accompanying Sharpe ratio in Fig.~\ref{fig:mabregret}. We observe that optimising $\alpha \to + 1^-; 2$ leads to the lowest cumulative regret and a high Sharpe ratio. Conversely, optimising $\alpha \to 0^+; \to + \infty$; leads to the highest cumulative regret and lowest Sharpe ratio. 

\begin{figure}[!htp]
    \centering
    \includegraphics[width=\textwidth]{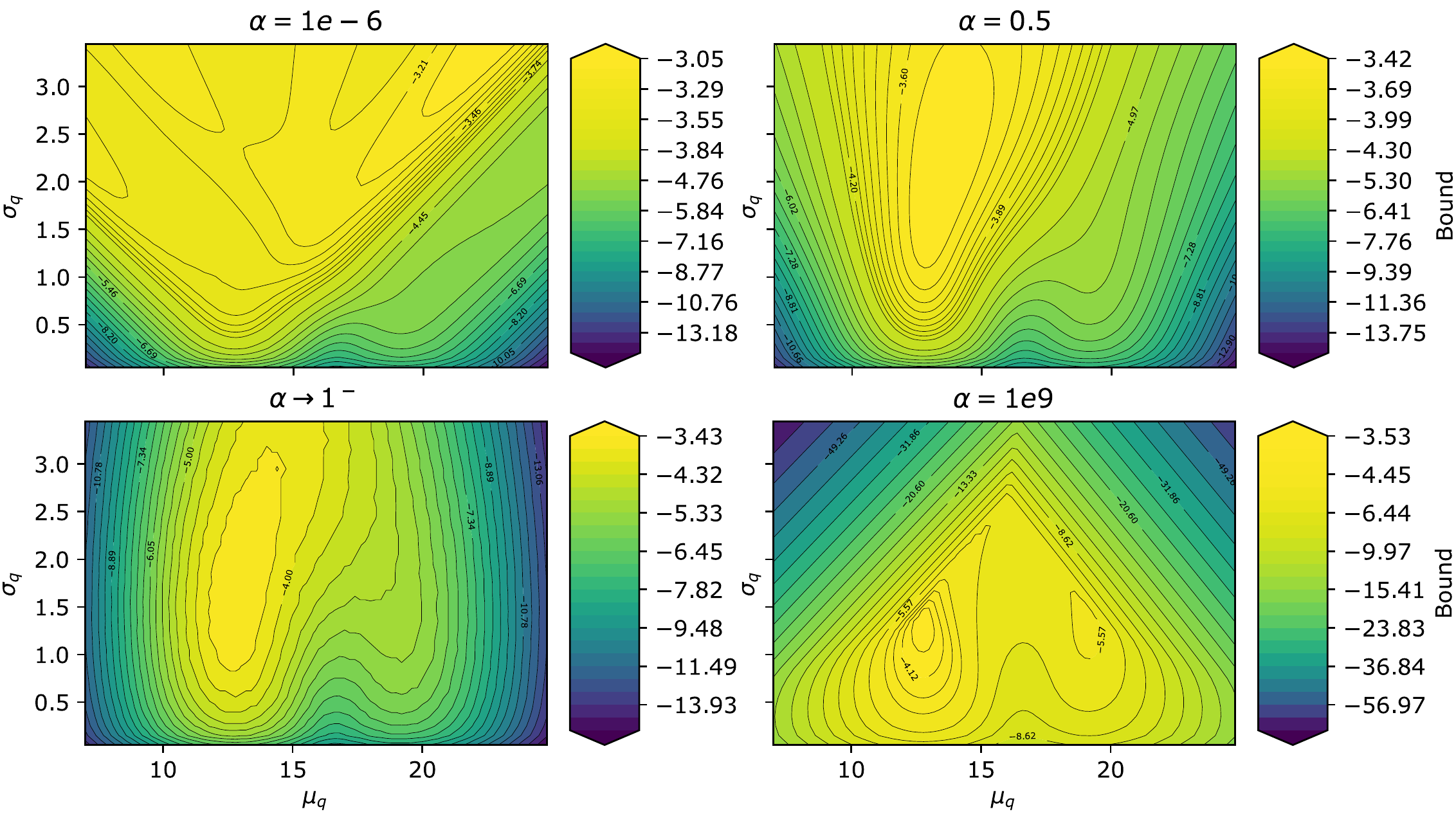}
    \caption{The R\'enyi bound as a function of the variational posterior. The contour plots show the optimisation landscape for each $\alpha$. For $\alpha = 1e9$ we observe two optima; for small $\alpha$ ($1e-6$) the optimal solution exhibits high variance.}
    \label{fig:cont1}
\end{figure}

To investigate this further, we plot the variational bounds for arm $1$ under different $\alpha$ parameters (Fig.~\ref{fig:cont1}). Recall from Fig.~\ref{fig:mabarm} that if the variational posterior fits the right-hand-side mode, this results in sub-optimal arm selection and the highest regret. This is because the agent would wrongly infer a high Sharpe ratio for this particular arm --- while it is in fact low --- increasing the probability that it was selected. We can explain the high regret of agents with $\alpha \to + \infty$; $\to 0^+$ from the property of their variational bound: For agents optimising $\alpha \to + \infty$, the approximate posterior fit the right-hand-side mode of the distribution due to its lower variance (i.e., mode-seeking behaviour). Conversely, agents with $\alpha \to 0^+$ would exhibit mass-covering, high-variance posterior estimates. In contrast, agents optimising $\alpha \to 1^-;0.5$ covered the left-hand-side mode and thus estimated a lower Sharpe ratio for this particular arm, which decreased the probability of it being selected (Fig~\ref{fig:cont1}).

These numerical experiments suggest that if agents sample their actions from posterior beliefs about what they are sampling, and those posterior beliefs depend on the form of the R\'enyi bound $\alpha$ parameterisation, then there is a natural space and explanation for behavioural variations. In short, the shape of the posterior –- that underwrites ensuing behaviour –- depends sensitively on the functional form of the variational bound.

\section{Discussion}\label{sec::discussion}
This paper accounts for behavioural variations among agents using R\'enyi divergences and their associated variational bounds. These divergences are R\'enyi relative entropies\footnote{The R\'enyi entropy provides a parametric family of measures of information (\cite{Renyi1961})}, and satisfy similar properties as the KL divergence (\cite{Renyi1961,Van2014}). R\'enyi divergences depend on an $\alpha$ parameter that controls the strength of the bound and induces different posterior estimates about the state of the world. In turn, different beliefs about the world lead to differences in behaviour. This provides a natural explanation as to why some people are more risk averse than others. For this alternative account to hold, we assumed throughout that agents sample their actions from posterior beliefs about the world, and those posterior beliefs depend on the form of the R\'enyi bound's $\alpha$ parameter. Yet, note that a similar account is possible if actions depended upon an expected free energy functional (\cite{friston2017process,parr2019generalised,vande2021,han2021goal}), intrinsic reward (\cite{Storck:95,schmidhuber2006curious,schmidhuber1991curious,sun2011}) or any class of objective functions that incorporates beliefs about the environment.

Crucially, R\'enyi divergences account for behavioural differences in a way that is formally distinct from a change in prior beliefs. This stems from the ability to disentangle different preference modes  by varying the bound's $\alpha$ parameter. Our simple multi-armed bandit setting illustrates this: e.g., large $\alpha$ values exhibit greater consistency in preferences. This contrasts with formal explanations based upon adjusting the precision or form of the prior under a variational bound based upon the KL divergence (i.e., $\alpha = 1$). Under active inference \cite{friston2017process,Costa20210}, multiple behavioural deficits have been illustrated by manipulation of the precision over the priors (\cite{parr2017uncertainty,RN18}). Although there has been some focus upon priors and on the form of the variational posterior (\cite{Schwobel2018}), relatively little attention has been paid to the nature of the bound itself in determining behaviour.

\subsection{Implications for the Bayesian brain hypothesis}

Our work is predicated on the idea that the brain is Bayesian and performs some sort of variational inference to infer its environment from its sensations. Practically, this entails the optimisation of a variational functional to make appropriate predictions. However, there are no unique functional forms for implementing such systems, and what variables account for differences in observed behaviour. On the basis of the above, we appeal to R\'enyi bounds, in addition to altered priors, to model behavioural variations. By committing to the R\'enyi bound, we provide an alternative perspective on how variant (or sub-optimal) behaviour can be modelled. This leads to a conceptual reversal of the standard variational free energy schemes, including predictive processing, etc \cite{bogaczTutorialFreeenergyFramework2017,buckleyFreeEnergyPrinciple2017}. That is, we can illustrate behavioural variations to be due to different variational objectives given particular priors, instead of different priors given the variational free energy. This has implications for how we model implementations of variational inference in the brain. That is, do we model sub-optimal inferences using altered generative models or alternative variational bounds? This turns out to be significant in light of our numerical analysis (section~\ref{sec::numanal}) that show no formal correspondence between these formulations. 

In a deep temporal system like the brain, one might ask if different cortical hierarchies might be performing inference under different variational objectives. It might be possible that variational objectives for lower levels to be modulated by higher levels through priors over $\alpha$ values -- a procedure of meta-inference. This is analogous to including precision priors over model parameters that have been associated with different neuromodulatory systems e.g., state transition precision with noradrenergic and sensory precision with cholinergic systems (\cite{fountas2020,parr2017uncertainty}). Consequently, this temporal separation of $\alpha$ parameterisations may provide an interesting research avenue for understanding the role of neuromodulatory systems and how they facilitate particular behaviours (\cite{Angela2002,Angela2005}).

\subsection{Generalised variational inference}
The R\'enyi bound provides a generalised variational inference objective derived from the R\'enyi divergence. This is because  R\'enyi divergences comprise the KL divergence as a special case (\cite{Minka2005}). These divergences allow us to naturally account for multiple behavioural preferences, directly via the optimisation objective, without changing prior beliefs. Other variational objectives can be derived from other general families of divergences such as f-divergences, Wasserstein distances (\cite{Ambrogioni2018,Dieng2016,Regli2018}), etc., which can improve the statistical properties of the variational bounds for particular applications (\cite{Wan2020,Zhang2019}). Future work could generalise the arguments presented here and examine how these different divergences shape behaviour when planning as inference.

\subsection{Limitations and future directions}
We do not observe a direct correspondence between the R\'enyi bound and the variational free energy under particular priors. However, our evaluations are based on a restricted Gaussian system. Therefore, future work should investigate this in more complex systems to show what sorts of prior modifications are critical in establishing similar optimisation landscapes for different variational bounds, in order to understand the relationship between the two. This will entail further exploring the association between the variational posterior and $\beta$ or $\alpha$ value.

Implementations of the R\'enyi bound are constrained by sampling biases and interesting differences in optimisation landscape. Indeed, when $\alpha$ is extremely large, even if the approximate posterior distribution belongs to the same family as the true posterior the optimisation becomes very difficult, causing the bound to be too conservative and introduce convergence issues. However, it must be noted that instances of this are due to the numerics of optimising the R\'enyi bound, rather than a failure of the bound itself. Practically, this means that careful consideration needs to be given to both the learning rate and stopping procedures during the optimisation of the R\'enyi bound. 

Our work includes implicit constraints on the form of the variational posterior. We have assumed a mean-field approximation in our simulations. However, this does not necessarily have to be the case. Interestingly, richer parameterisations of the variational posterior might negate the impact of the $\alpha$ values. Specifically, we noted that if the true posterior is in the same family of distributions as the variational posterior, then changing the $\alpha$ value does not impact the shape of the variational posterior --- and consequently the system’s behaviour. However, complex parameterisations are computationally expensive and can still be inappropriate. Therefore, this departure from vanilla variational inference provides a useful explanation for different behaviours that biological (or artificial) agents might adopt --- under the assumption that the brain performs variational Bayesian inference. Orthogonal to this, an interesting future direction is investigating the connections between the variational posterior form and how it may impact the variational bound. This has direct consequences for the types of message passing schemes that might be implemented in the brain (\cite{Minka2005,parr2019message}).

\section{Conclusion}\label{sec::conclusion}
We offer an account of behavioural variations using  R\'enyi divergences and their associated variational bounds bounds that complement usual formulations in terms of different prior beliefs. We show how different R\'enyi bounds induce behavioural differences for a fixed generative model that are formally distinct from a change of priors. This is accomplished by changes in an $\alpha$  parameter that alters the bound’s strength, inducing different inferences and consequent behavioural variations. Crucially, the inferences produced in this way do not seem to be accounted for by a change in priors under the standard variational objective. We emphasise that the R\'enyi bounds are analogous to the variational free energy (or evidence lower bound) and can be derived using the same assumptions. This formulation is illustrated through numerical analysis and demonstrates that $\alpha >1$ values give rise to mode-seeking behaviours and $\alpha < 1$ values to mode-covering behaviours when priors are held constant.

\paragraph*{Software note} The code required to reproduce the simulations and figures is available here:

\href{https://github.com/ucbtns/renyibounds}{https://github.com/ucbtns/renyibounds}

\paragraph*{Acknowledgements} NS is funded by Medical Research Council (MR/S502522/1). FF is funded by the ERC Advanced Grant (no: 742870) and by the Swiss National Supercomputing Centre (CSCS, project: s1090). LD is supported by the Fonds National de la Recherche, Luxembourg (Project code: 13568875). This publication is based on work partially supported by the EPSRC Centre for Doctoral Training in
Mathematics of Random Systems: Analysis, Modelling and Simulation (EP/S023925/1). KJF is funded by the Wellcome Trust (Ref: 203147/Z/16/Z and 205103/Z/16/Z). 

\paragraph{Conflicts of Interest} The authors declare no conflict of interest 

\bibliography{paper}  
\newpage
\appendix
\section{Note on divergences indexed by $\alpha$}\label{appendix::divergences}
Many divergences in the literature are indexed with the parameter $\alpha$ (see Table~\ref{table:alphas}). These divergences turn out to be equivalent to the R\'enyi divergence as we can identify one-to-one correspondences between them.

\begin{table}[h!]
 \footnotesize
\centering
\begin{tabular}{c || c}
 \hline
Divergence & Formulation  \\ [0.5ex] 
 \hline\hline
I-divergence (\cite{Nielsen2011}) & $D_{\alpha}^{I}[p||q]= \int_{\mathcal{S}} p^{\alpha} q^{(1-\alpha)} ds$\\ 
Amari's $\alpha$ divergence (\cite{Amari2009alphaI}) & $D^{AM}_{\alpha} [p||q]=\frac{4}{1-\alpha^{2}} (1-\int_{\mathcal{S}} p^{\frac{1+\alpha}{2}}  q^{\frac{1-\alpha}{2}} ds)$\\
Tsallis' divergence (\cite{Nielsen2011}) & $ D^{T}_{\alpha} [p||q]= \frac{1}{\alpha-1} (\int_{\mathcal{S}} p^{\alpha}  q^{1-\alpha} ds-1)$\\
 R\'enyi divergence &   $D_{\alpha}[p||q]= \frac{1}{(\alpha -1)} \log \int_{\mathcal{S}} p^{\alpha} q^{(1-\alpha)}\de s$\\ [1ex] 
\hline
\end{tabular}
\caption{Divergence families indexed with $\alpha$. Amari’s $\alpha$-divergence plays an important role in information geometry as it induces a dually-flat geometry on the space of probability measures, and furthermore, when extended to positive measures, it is the only intersection between f-divergences and Bregman divergences, two important families of divergences (\cite{Amari2009alphaI,Ay2016}). }
\label{table:alphas}
\end{table}
\normalsize

All of the divergences shown in Table~\ref{table:alphas} are equivalent, in the sense that there are one-to-one mappings between them.

The Tsallis and Amari’s divergences are linear functions of the I-divergence:

\begin{align}
    D^{T}_{\alpha}[p||q] &= \frac{1}{\alpha-1} (D_{\alpha}^{I}[p||q] -1) \\
    D^{AM}_{\alpha}[p||q] &= \frac{4}{1-\alpha^{2}} (1-D_{\frac{1+\alpha}{2}}^{I}[p||q])
\end{align}

As a consequence, the Amari  $\alpha$ divergence is a scalar multiple of the Tsallis divergence, under the correspondence $ \beta = \frac{1+\alpha}{2}$:

\begin{align}
    D^{AM}_{\alpha}[p||q] &= \frac{1}{\beta} D^{T}_{\beta}[p||q] 
\end{align}

Finally, the R\'enyi divergence is a monotonic function of the I-divergence:

\begin{align}
    D_{\alpha}[p||q] &= \frac{1}{\alpha -1} \log   D^{I}_{\alpha}[p||q] 
\end{align}

\section{Derivations}\label{appendix::derivations}
\subsection{Negative variational free energy for Gaussian-Gamma distribution}
Here, we work through the variational free energy for the system described in Section~\ref{sec::post}. $s,o$ are the random variables of interest, $x$ the parameter governing the mean and $\Sigma_k$ is the precision parameter:

\begin{align}
    p(s|\lambda_p) & \sim \mathcal{N}(s;0,(\lambda_p \sigma_p)^{-1})Gam(\lambda_p ;\alpha_p,\beta_p) \\
   p(o|s) & \sim \mathcal{N}(o;sx, \Sigma_l) \\
   q(s) & \sim \mathcal{N}(s;\mu_q,\Sigma_q)
\end{align}

where $\Sigma_k=(\lambda_k \sigma_k )^{-1}$,
The probability density functions are defined as:

\small
\begin{align}
    p(s|\lambda_p) &= \frac{|\lambda_p \sigma_p|^{1/2}}{2\pi^{1/2}} \exp \Big[\frac{-\lambda_p}{2} s^T \sigma_p s \Big] \frac{\beta_p^{\alpha_p}}{\Gamma(\alpha_p)} \lambda_p^{\alpha_p-1} \exp \Big[-\lambda_p \beta_p \Big]\\
    p(o|s) &= \frac{|\Sigma_l|^{-1/2}}{2\pi^{n/2}} \exp \Big[-\frac{1}{2} (o-sx)^T \Sigma_l^{-1} (o-sx) \Big] \\
    q(s) &= \frac{|\Sigma_q|^{-1/2}}{2\pi^{1/2}} \exp \Big[-\frac{1}{2} (s-\mu_q)^T \Sigma_q^{-1} (s-\mu_q) \Big]
\end{align}
\normalsize

We use probability distributions to derive the quantity of interest: $\mathbb{E}_{q(s)}[\log p(s,o) - \log q(s) ] = -D_{KL}[q(s)||p(s,o)] $:

\scriptsize
\begin{align}
    -D_{KL}&[q(s)||p(s,o)] = - \int_{\mathcal{S}}q(s) \log \left( \frac{q(s)}{p(s,o)}  \right) \de s =\\
    & = - \int_{\mathcal{S}}q(s) \log \left[ \frac{(2\pi)^{\frac{n+1}{2}}|\Sigma_p|^{1/2}|\Sigma_l|^{1/2}}{(2\pi)^{1/2}|\Sigma_q|^{1/2}} \frac{\beta_p^{\alpha_p} \lambda_p^{\alpha_{p-1}} \exp(-\lambda_p \beta_p)}{\Gamma(\alpha_p)} \right] \de s +\\ & +\int_{\mathcal{S}}q(s) \left [ \frac{1}{2}(s-\mu_q)^2 \Sigma_q^{-1} - \frac{1}{2}\left( (o-sx)^T\Sigma_l^{-1}(o-sx) + s^T\Sigma_p^{-1}s \right)  \right ] \de s = \\
    & = \frac{1}{2}  \log \left[  \frac{|\Sigma_q|}{(2\pi)^n|\Sigma_p||\Sigma_l|} \right] + \log \left [ \frac{\beta_p^{\alpha_p} \lambda_p^{\alpha_{p-1}}}{\Gamma(\alpha_p) }\right]-\lambda_p \beta_p +\\
    & + \int_{\mathcal{S}}q(s) \left [ -\frac{1}{2}[s^2(\Sigma_p^{-1} + x^T \Sigma_l^{-1}x - \Sigma_q^{-1}) -2s(-\mu_q\Sigma_q^{-1} + x^T \Sigma_l^{-1}o) -\mu_q^2\Sigma_q^{-1} + o^T \Sigma_l^{-1}o] \right ] \de s \label{eq:kl_def}
    \end{align}
\normalsize

Consider the last integral:
  
    \scriptsize
    \begin{align}
        -\frac{1}{2}(\Sigma_p^{-1} & + x^T \Sigma_l^{-1}x - \Sigma_q^{-1}) \int_{\mathcal{S}} s^2 q(s)\de s +(-\mu_q\Sigma_q^{-1} + x^T \Sigma_l^{-1}o) \int_{\mathcal{S}} s q(s)\de s -\frac{1}{2} (-\mu_q^2\Sigma_q^{-1} + o^T \Sigma_l^{-1}o) \int_{\mathcal{S}} q(s)\de s  =\\
        & -\frac{1}{2}(\Sigma_p^{-1} + x^T \Sigma_l^{-1}x - \Sigma_q^{-1})(\Sigma_q + \mu_q^2)  +(-\mu_q\Sigma_q^{-1} + x^T \Sigma_l^{-1}o) \mu_q -\frac{1}{2} (-\mu_q^2\Sigma_q^{-1} + o^T \Sigma_l^{-1}o) =\\
          & - \frac{1}{2}(\Sigma_q\Sigma_p^{-1} + \Sigma_qx^T \Sigma_l^{-1}x - 1 + \mu_q^2\Sigma_p^{-1} + \mu_q^2x^T \Sigma_l^{-1}x -2 \mu_q x^T \Sigma_l^{-1}o + o^T \Sigma_l^{-1}o)     
    \end{align}
\normalsize

Combining the results we have:

\begin{align}  
-D_{KL}[q(s)||p(s,o)] & = \frac{1}{2}\log\left(\frac{|\Sigma_q|}{(2\pi)^n |\Sigma_p| |\Sigma_l|}\right) \\
    & - \frac{1}{2}\left(o^T\Sigma_l^{-1}o + \mu_q^2 \Sigma_p^{-1} + \mu_q^2 x^T \Sigma_l^{-1} x -2\mu_q x^T \Sigma_l^{-1} o\right)  \\
    & -\frac{1}{2}\left(\Sigma_qx^T \Sigma_l^{-1} x + \Sigma_q\Sigma_p^{-1} - 1 \right) \\
    & + \log \left [ \frac{\beta_p^{\alpha_p} \lambda_p^{\alpha_{p-1}}}{\Gamma(\alpha_p)}\right]-\lambda_p \beta_p
\end{align}

\subsection{R\'enyi bound for Gaussian distribution}

The probability density function for the random variables, $s, o$ and $x$ is the parameter governing the means:

\begin{align}
    p(s)& =\frac{1}{2\pi^{\frac{1}{2}} |\Sigma_p|^{\frac{1}{2}}}  \exp\big[-\frac{1}{2} s^T\Sigma^{-1}_{p} s\big]\\
    p(o|s)& =\frac{1}{2\pi^{\frac{n}{2}} |\Sigma_l|^{\frac{1}{2}}}  \exp \big[-\frac{1}{2} (o-sx)^T\Sigma^{-1}_{l} (o-sx) \big]\\
    q(s)& =\frac{1}{2\pi^{\frac{1}{2}} |\Sigma_q|^{\frac{1}{2}}}  \exp\big[(-\frac{1}{2} (s-\mu_q)^T\Sigma^{-1}_{q} (s-\mu_q)\big]
\end{align}

We now supplement these quantities into the negative R\'enyi bound, and rewrite using the defined quantities:

\begin{align}
    -D_{\alpha} &[q(s)||p(s,o)] = \frac{1}{1-\alpha} \log \int_{\mathcal{S}} q(s)^{\alpha} p(s,o)^{1-\alpha} \de s \\
    & = \frac{1}{1-\alpha} \log \left( \frac{1}{2\pi^{\alpha /2}
    2\pi^{(1-\alpha)\frac{n+1}{2}} |\Sigma_q|^{\alpha/2}|\Sigma_p|^{(1-\alpha)\frac{1}{2}}|\Sigma_l|^{(1-\alpha)\frac{1}{2}}} \right) \\
    & + \frac{1}{1-\alpha}\log \int_{\mathcal{S}} \exp \Bigg( -\frac{1}{2}\Bigg[\alpha [(s-\mu_q)^T\Sigma^{-1}_{q} (s-\mu_q)] + \\ 
    & (1-\alpha)[o^T\Sigma^{-1}_{l}o - 2s^T x^T \Sigma^{-1}_{l}o +
     s^T[\Sigma^{-1}_{p} +x^T \Sigma^{-1}_{l}x ]s] \Bigg] \Bigg)\de s\\
    & = \frac{1}{2} \log \left( \frac{\Sigma_q^{\frac{\alpha}{\alpha-1}}}{2\pi^{\frac{n(1-\alpha)-1}{1-\alpha}}
    |\Sigma_p||\Sigma_l|} \right)\\
    & - \frac{1}{\alpha-1} \log \int_{\mathcal{S}} \exp \Bigg( -\frac{1}{2} \Bigg[ s^T(\alpha \Sigma_q^{-1} + x^T \Sigma^{-1}_l x(1-\alpha) + \Sigma_p^{-1}(1-\alpha))s \label{eq:ren1}\\
    &- 2s((1-\alpha) x^T \Sigma^{-1}_l o +\alpha \mu_q \Sigma_q^{-1}) + \alpha \mu_q^2 \Sigma^{-1}_q + (1-\alpha) o^T \Sigma_l^{-1}o \Bigg] \Bigg) \de s \label{eq:ren2}
\end{align}

First, let us focus on the term inside the integral. To avoid clutter we replace: $\Sigma_{\alpha}^{-1} := \alpha \Sigma_q^{-1} + x^T \Sigma_l^{-1}x(1-\alpha) + \Sigma_p^{-1}(1-\alpha)$ and assume it is invertible. We define $\mu_{\alpha} := \Sigma_{\alpha}(\alpha \mu_q \Sigma_q^{-1} + (1-\alpha) x^T \Sigma_l^{-1}o)$. Then Eq.~\ref{eq:ren1} and~\ref{eq:ren2} can be rewritten as:

\begin{align}
    & -\frac{1}{2}\frac{-(1-\alpha)o^T \Sigma_l^{-1}o - \alpha \mu_q^2 \Sigma_q^{-1} + \mu_{\alpha}^2 \Sigma_{\alpha}^{-1}}{\alpha-1} \\
    & -\frac{1}{\alpha-1} \log \int_{\mathcal{S}} \frac{2 \pi ^{1/2} |\Sigma_{\alpha}|^{1/2}}{2 \pi ^{1/2} |\Sigma_{\alpha}|^{1/2}} \exp \Bigg( -\frac{1}{2} (s-\mu_{\alpha})^T \Sigma_{\alpha}^{-1}(s-\mu_{\alpha})\Bigg) \de s \\
    & = -\frac{1}{2(1- \alpha)} \Bigg[(1-\alpha)o^T \Sigma_l^{-1}o + \alpha\mu_q^2 \Sigma_q^{-1} - \mu_{\alpha}^2 \Sigma_{\alpha}^{-1} \Bigg] + \frac{1}{2}\log ( 2\pi^{\frac{1}{1-\alpha}} \Sigma_{\alpha}^{\frac{1}{1-\alpha}})
\end{align}

Putting it all together:

\begin{align}
    D_{\alpha}[q(s)||p(s,o)]  &= \frac{1}{2} \log \left( \frac{\Sigma_q^{\frac{\alpha}{\alpha-1}} |\Sigma_{\alpha}^{-1}|^{\frac{1}{\alpha-1}}}{2\pi^n
    |\Sigma_p||\Sigma_l|} \right) \\
    & -\frac{1}{2} \Bigg[o^T \Sigma_l^{-1}o - \frac{\alpha}{(\alpha-1)}\mu_q^2 \Sigma_q^{-1} + \frac{1}{(\alpha-1)} \mu_{\alpha}^2 \Sigma_{\alpha}^{-1}\Bigg]
\end{align}

With this formulation, we turn to the first term:

\begin{align}
&\frac{1}{2} \log \left[ \frac{|\Sigma_q|^{\frac{\alpha}{\alpha-1}}|\Sigma_{\alpha}^{-1}|^{\frac{1}{\alpha-1}}}{(2\pi)^{n}|\Sigma_p| |\Sigma_l|} \right] \\
&=\frac{1}{2} \log \left[ \frac{|\Sigma_q|}{(2\pi)^{n}|\Sigma_p| |\Sigma_l|} \right] - \frac{1}{2} \log (\Sigma_q \Sigma_{\alpha}^{-1})^{\frac{1}{1-\alpha}} \\
&=\frac{1}{2} \log \left[ \frac{|\Sigma_q|}{(2\pi)^{n}|\Sigma_p| |\Sigma_l|} \right]\\ 
&- \frac{1}{2(1-\alpha)} \log \Bigg(1 +(1-\alpha)(\Sigma_qx^T \Sigma_l^{-1} x + \Sigma_q\Sigma_p^{-1} - 1)    \Bigg)
\end{align}

Now, let us consider the second term:

\small
\begin{align}
& -\frac{1}{2} \Bigg[o^T \Sigma_l^{-1}o - \frac{\alpha}{(\alpha-1)}\mu_q^2 \Sigma_q^{-1} + \frac{1}{(\alpha-1)} \mu_{\alpha}^2 \Sigma_{\alpha}^{-1}\Bigg]\\
&= -\frac{1}{2} \Bigg[o^T\Sigma_l^{-1}o  - \frac{\alpha}{(\alpha-1)} \mu_q^2 \Sigma_q^{-1} \\
&+ \frac{1}{(\alpha-1)}\frac{\alpha^2 \mu_q^2 (\Sigma_q^{-1})^2 + (1-\alpha)^2(x^T \Sigma_l^{-1}o)^2 +2\alpha (1-\alpha)\mu_q \Sigma_q^{-1}x^T \Sigma_l^{-1}o}{\Sigma_{\alpha}^{-1}}   \Bigg]\\
&= -\frac{1}{2} \Bigg[o^T\Sigma_l^{-1}o +\frac{-\alpha^2 \mu_q^2 (\Sigma_q^{-1})^2 - (1-\alpha)\alpha \mu_q^2 \Sigma_q^{-1}\Sigma_p^{-1}} {(\alpha-1)( \alpha \Sigma_q^{-1} + (1-\alpha)( \Sigma_p^{-1} + x^T\Sigma_l^{-1}x))} \\
&- \frac{(1-\alpha)\alpha \mu_q^2 \Sigma_q^{-1}x^T \Sigma_l^{-1}x  +\alpha^2 \mu_q^2 (\Sigma_q^{-1})^2 + (1-\alpha)^2(x^T \Sigma_l^{-1}o)^2 }{(\alpha-1)( \alpha \Sigma_q^{-1} + (1-\alpha)( \Sigma_p^{-1} + x^T\Sigma_l^{-1}x))} \\
&+ \frac{2\alpha (1-\alpha)\mu_q \Sigma_q^{-1}x^T \Sigma_l^{-1}o}{(\alpha-1)( \alpha \Sigma_q^{-1} + (1-\alpha)( \Sigma_p^{-1} + x^T\Sigma_l^{-1}x))}\Bigg]\\
&= -\frac{1}{2} \Bigg[o^T\Sigma_l^{-1}o +\frac{\alpha \mu_q^2 \Sigma_q^{-1}\Sigma_p^{-1} +\alpha \mu_q^2 \Sigma_q^{-1}x^T \Sigma_l^{-1}x}{\alpha \Sigma_q^{-1} + (1-\alpha)( \Sigma_p^{-1} + x^T\Sigma_l^{-1}x)}   \\
&+ \frac{ -(1-\alpha)(x^T \Sigma_l^{-1}o)^2 -2\alpha \mu_q \Sigma_q^{-1}x^T \Sigma_l^{-1}o}{\alpha \Sigma_q^{-1} + (1-\alpha)( \Sigma_p^{-1} + x^T\Sigma_l^{-1}x)}\Bigg]\\
&= -\frac{1}{2} \Bigg[\frac{\alpha \Sigma_q^{-1}o^T\Sigma_l^{-1}o +(1-\alpha) \Sigma_p^{-1} o^T\Sigma_l^{-1}o + (1-\alpha)o^T\Sigma_l^{-1}o x^T\Sigma_l^{-1}x}{\alpha \Sigma_q^{-1} + (1-\alpha)( \Sigma_p^{-1} + x^T\Sigma_l^{-1}x)}   \\
&+ \frac{\alpha \mu_q^2 \Sigma_q^{-1}\Sigma_p^{-1} +\alpha \mu_q^2 \Sigma_q^{-1}x^T \Sigma_l^{-1}x - (1-\alpha)(x^T \Sigma_l^{-1}o)^2 -2\alpha \mu_q \Sigma_q^{-1}x^T \Sigma_l^{-1}o}{\alpha \Sigma_q^{-1} + (1-\alpha)( \Sigma_p^{-1} + x^T\Sigma_l^{-1}x)} \Bigg]\\
&= -\frac{1}{2} \Bigg[\frac{\alpha \Sigma_q^{-1}o^T\Sigma_l^{-1}o +(1-\alpha) \Sigma_p^{-1} o^T\Sigma_l^{-1}o +\alpha \mu_q^2 \Sigma_q^{-1}\Sigma_p^{-1}}{\alpha \Sigma_q^{-1} + (1-\alpha)( \Sigma_p^{-1} + x^T\Sigma_l^{-1}x)}   \\
&+ \frac{\alpha \mu_q^2 \Sigma_q^{-1}x^T \Sigma_l^{-1}x -2\alpha \mu_q \Sigma_q^{-1}x^T \Sigma_l^{-1}o}{\alpha \Sigma_q^{-1} + (1-\alpha)( \Sigma_p^{-1} + x^T\Sigma_l^{-1}x)} \Bigg]\\
&= -\frac{\alpha}{2 \Sigma_q \Sigma_{\alpha}^{-1}} \Bigg[o^T\Sigma_l^{-1}o +\Sigma_q \frac{(1-\alpha)}{\alpha} \Sigma_p^{-1} o^T\Sigma_l^{-1}o +\mu_q^2\Sigma_p^{-1} \\
&+\mu_q^2 x^T \Sigma_l^{-1}x -2\mu_q x^T \Sigma_l^{-1}o\Bigg]
\end{align}
\normalsize

From this, the simplified formulation for the R\'enyi bound is:

\begin{align}
D_{\alpha}[q(s)||p(s,o)] & = \frac{1}{2}\log\left(\frac{|\Sigma_q|}{(2\pi)^n |\Sigma_p| |\Sigma_l|}\right) \\
    &- \frac{\alpha}{2(\Sigma_q \Sigma_{\alpha}^{-1})}\left(o^T\Sigma_l^{-1}o + \mu_q^2 \Sigma_p^{-1} + \mu_q^2 x^T \Sigma_l^{-1} x -2\mu_q x^T \Sigma_l^{-1} o\right)\\
    &- \frac{1}{2(1-\alpha)} \log \left(1+(1-\alpha)(\Sigma_qx^T \Sigma_l^{-1} x + \Sigma_q\Sigma_p^{-1} - 1)    \right) \\
    &- \frac{1}{2 \Sigma_{\alpha}^{-1}}\left((1-\alpha)\Sigma_p^{-1} o^T \Sigma_l^{-1}o \right)
\end{align}

\section{MAB experiment details}\label{appendix::code}

We implemented the MAB simulations as described in Algorithm~\ref{alg:rb}. 

\begin{algorithm}[H]
  \caption{MAB optimisation using R\'enyi variational inference}
  \label{alg:rb}
   \hspace*{\algorithmicindent} \textbf{Input}: Variational density $q(s)$ for each arm. Empty observation buffer $D_i$ for each arm $i$\\
   \hspace*{\algorithmicindent} \textbf{Output} : Optimal arm selection
    \begin{algorithmic}
    \State Initialise $\mu_q^i, \Sigma_q^i$ for each arm $i$
	\Repeat:
		\For {each arm i}:
    		\State Sample $s_i \sim q(\cdot|\mu_q^i, \Sigma_q^i)$
        \EndFor
        \State Compute $i^* = \argmax_i \frac{s_i}{\Sigma_q^i}$
        \State Pull arm $i^*$, receive reward $R_i^*$ and store it in $D_{i^*}$
		\For {each arm i}:
		    \State Update variational parameters by gradient descent:
		    \State $\nabla_{\mu_q^i} D_{\alpha}(q(s)|| p(s,o))$
		    \State $\nabla_{\Sigma_q^i} D_{\alpha}(q(s)|| p(s,o))$

        \EndFor
    \Until{convergence}
    \end{algorithmic}
\end{algorithm}
For learning, the experiments were parameterised as:
\begin{itemize}
    \item $4000$ iterations for each simulation.
    \item R\'enyi bound was optimised using ADAM with a learning rate of $2e-2$ and $10$ updates.
    \item $300$ Monte-Carlo samples were used to compute the variational posterior, $q(s)$ at each iteration
\end{itemize}

For each simulation, the prior specification is shown in Table~\ref{table::mabspec}, and generative process in Table~\ref{table::mabprocess}.

\begin{table}[!ht]
    \centering
    \begin{tabular}{|c||c|c|c|}
        \hline
         &  $\mu$ & $\Sigma$ & Weights \\
         \hline  
          $q(s)$: & $25$ & $1e-8$ & $\cdot$\\
         Arm $1$  & $13,20$ & $1.5,1.5$ & $0.5$\\
         Arm $2$ & $16,14$ & $1.5,1.5$ & $0.5$\\
         Arm $3$ & $10,17$ & $1.5,1.5$ & $0.5$\\
         \hline
    \end{tabular}
        \caption{MAB multi-modal priors.}
    \label{table::mabspec}
\end{table}

\begin{table}[!ht]
    \centering
    \begin{tabular}{|c|| c|c|c|}
        \hline
         & $\mu$ & $\Sigma$ & Weights \\
         \hline  
         Arm $1$ & $10,22$ & $1,1$ & $0.97,0.3$\\
         Arm $2$ & $16$ & $3$ & $\cdot$\\
         Arm $3$ & $10,10$ & $1,1$ & $0.97,0.03$\\
         \hline
    \end{tabular}
        \caption{MAB multi-modal generative process.}
    \label{table::mabprocess}
\end{table}

\end{document}


\maketitle

\section{KL bound}

\begin{align*}
    -D_{KL}(q(s)||p(s,o)) & = - \int_{\mathcal{S}}q(s) \log \left( \frac{q(s)}{p(s,o)}  \right) \de s =\\
    & = - \int_{\mathcal{S}}q(s) \log \left[ \frac{(2\pi)^{\frac{n+1}{2}}|\Sigma_p|^{1/2}|\Sigma_l|^{1/2}}{(2\pi)^{1/2}|\Sigma_q|^{1/2}} \right] \de s +\\ & +\int_{\mathcal{S}}q(s) \left [ \frac{1}{2}(s-\mu_q)^2/\Sigma_q - \frac{1}{2}\left( (o-sx)^T\Sigma_l^{-1}(o-sx) \right)  \right ] \de s = \\
    & = \frac{1}{2}  \log \left[  \frac{|\Sigma_q|}{(2\pi)^n|\Sigma_p||\Sigma_l|} \right] +\\
    & + \int_{\mathcal{S}}q(s) \left [ -\frac{1}{2}[s^2(\Sigma_p^{-1} + x^T \Sigma_l^{-1}x - \Sigma_q^{-1}) -2s(-\mu_q\Sigma_q^{-1} + x^T \Sigma_l^{-1}o) -\mu_q^2\Sigma_q^{-1} + o^T \Sigma_l^{-1}o] \right ] \de s = \\
    & = \frac{1}{2}  \log \left[  \frac{|\Sigma_q|}{(2\pi)^n|\Sigma_p||\Sigma_l|} \right] +\\
    & + \int_{\mathcal{S}}q(s) \left [ -\frac{1}{2}[(s-\mu_{KL})^2\Sigma_{KL}^{-1} -\mu_{KL}^2\Sigma_{KL}^{-1}-\mu_q^2\Sigma_q^{-1} + o^T \Sigma_l^{-1}o] \right ] \de s = \\
    & = \frac{1}{2}  \log \left[  \frac{|\Sigma_q|}{(2\pi)^n|\Sigma_p||\Sigma_l|} \right] -\\
    &  \frac{1}{2} ( -\mu_{KL}^2\Sigma_{KL}^{-1}-\mu_q^2\Sigma_q^{-1} + o^T \Sigma_l^{-1}o) - \\ & \frac{1}{2}\Sigma_{KL}^{-1}\left[\int_{\mathcal{S}}s^2 q(s)\de s + \int_{\mathcal{S}}\mu_{KL}^2q(s) \de s  -2\mu_{KL} \int_{\mathcal{S}}s q(s) \de s \right] = \\
    & = \frac{1}{2}  \log \left[  \frac{|\Sigma_q|}{(2\pi)^n|\Sigma_p||\Sigma_l|} \right] -\\
    &  \frac{1}{2} ( -\mu_{KL}^2\Sigma_{KL}^{-1}-\mu_q^2\Sigma_q^{-1} + o^T \Sigma_l^{-1}o +\Sigma_q \Sigma_{KL}^{-1} + \mu_q^2 \Sigma_{KL}^{-1}+ \mu_{KL}^2\Sigma_{KL}^{-1} - 2 \mu_q \mu_{KL}\Sigma_{KL}^{-1}) =\\
    & = \frac{1}{2}  \log \left[  \frac{|\Sigma_q|}{(2\pi)^n|\Sigma_p||\Sigma_l|} \right] -\\
      \frac{1}{2} ( -\mu_q^2\Sigma_q^{-1} +& o^T \Sigma_l^{-1}o +\Sigma_q (\Sigma_p^{-1} + x^T \Sigma_l^{-1}x - \Sigma_q^{-1}) - 2 \mu_q x^T \Sigma_l^{-1}o  + 2 \mu_q^2 \Sigma_q^{-1} +\mu_q^2 x^T \Sigma_l^{-1}x + \mu_q^2\Sigma_p^{-1} -\mu_q^2 \Sigma_q^{-1} ) =\\    
        & \frac{1}{2}\log\left(\frac{|\Sigma_q|}{(2\pi)^n |\Sigma_p| |\Sigma_l|}\right) - \\
    & \frac{1}{2}\left(o^T\Sigma_l^{-1}o - \mu_q^2 \Sigma_p^{-1} + \mu_q^2 x^T \Sigma_l^{-1} x -2\mu_q x^T \Sigma_l^{-1} o\right) - \\
    & \frac{1}{2}\left(\Sigma_qx^T \Sigma_l^{-1} x + \Sigma_q\Sigma_p^{-1} - 1 \right)
\end{align*}

\section{Renyi bound}

\begin{align*}
    D_{\alpha}(q(s)||p(s,o)) & = \dots \\
    & = \frac{1}{2(1-\alpha)} \log \left[ \frac{|\Sigma_{\alpha}|}{(2\pi)^{(1-\alpha)n}|\Sigma_q|^{\alpha} |\Sigma_p|^{1-\alpha} |\Sigma_l|^{1-\alpha}} \right] -\\
    & \frac{1}{2(1-\alpha)} \left[ -\mu_{\alpha}^2\Sigma_{\alpha}^{-1} + \alpha \mu_q^2 \Sigma_q^{-1} + (1-\alpha) o^T\Sigma_l^{-1}o   \right]
\end{align*}
Let's focus on the first term:
\begin{align*}
&\frac{1}{2(1-\alpha)} \log \left[ \frac{|\Sigma_{\alpha}|}{(2\pi)^{(1-\alpha)n}|\Sigma_q|^{\alpha} |\Sigma_p|^{1-\alpha} |\Sigma_l|^{1-\alpha}} \right]  =\\
&\frac{1}{2} \log \left[ \frac{|\Sigma_{\alpha}|^{\frac{1}{1-\alpha}}}{(2\pi)^{n}|\Sigma_q|^{\frac{\alpha}{1-\alpha}}|\Sigma_p| |\Sigma_l|} \right]  =\\
&\frac{1}{2} \log \left[ \frac{|\Sigma_q|^{\frac{\alpha}{\alpha-1}}|\Sigma_{\alpha}^{-1}|^{\frac{1}{\alpha-1}}}{(2\pi)^{n}|\Sigma_p| |\Sigma_l|} \right]  =\\
&\frac{1}{2} \log \left[ \frac{|\Sigma_q|}{(2\pi)^{n}|\Sigma_p| |\Sigma_l|} \right] + \frac{1}{2} \log (\Sigma_q \Sigma_{\alpha}^{-1})^{\frac{1}{1-\alpha}}  =\\
&\frac{1}{2} \log \left[ \frac{|\Sigma_q|}{(2\pi)^{n}|\Sigma_p| |\Sigma_l|} \right] + \frac{1}{2(1-\alpha)} \log \left(1+(1-\alpha)(\Sigma_qx^T \Sigma_l^{-1} x + \Sigma_q\Sigma_p^{-1} - 1)    \right)\\
\end{align*}
Let's consider now the second term:
\tiny
\begin{align*}
& \frac{1}{2(1-\alpha)} \left[ -\mu_{\alpha}^2\Sigma_{\alpha}^{-1} + \alpha \mu_q^2 \Sigma_q^{-1} + (1-\alpha) o^T\Sigma_l^{-1}o   \right]=\\
& \frac{1}{2} \left[o^T\Sigma_l^{-1}o  - \frac{\alpha}{(\alpha-1)} \mu_q^2 \Sigma_q^{-1} + \frac{1}{(\alpha-1)}\mu_{\alpha}^2\Sigma_{\alpha}^{-1}    \right]=\\
& \frac{1}{2} \left[o^T\Sigma_l^{-1}o  - \frac{\alpha}{(\alpha-1)} \mu_q^2 \Sigma_q^{-1} + \frac{1}{(\alpha-1)}\frac{\alpha^2 \mu_q^2 (\Sigma_q^{-1})^2 + (1-\alpha)^2(x^T \Sigma_l^{-1}o)^2 +2\alpha (1-\alpha)\mu_q \Sigma_q^{-1}x^T \Sigma_l^{-1}o}{\Sigma_{\alpha}^{-1}}   \right]=\\
& \frac{1}{2} \left[o^T\Sigma_l^{-1}o +\frac{-\alpha^2 \mu_q^2 (\Sigma_q^{-1})^2 - (1-\alpha)\alpha \mu_q^2 \Sigma_q^{-1}\Sigma_p^{-1} - (1-\alpha)\alpha \mu_q^2 \Sigma_q^{-1}x^T \Sigma_l^{-1}x  +\alpha^2 \mu_q^2 (\Sigma_q^{-1})^2 + (1-\alpha)^2(x^T \Sigma_l^{-1}o)^2 +2\alpha (1-\alpha)\mu_q \Sigma_q^{-1}x^T \Sigma_l^{-1}o}{(\alpha-1)( \alpha \Sigma_q^{-1} + (1-\alpha)( \Sigma_p^{-1} + x^T\Sigma_l^{-1}x))}   \right]=\\
& \frac{1}{2} \left[o^T\Sigma_l^{-1}o +\frac{ +\alpha \mu_q^2 \Sigma_q^{-1}\Sigma_p^{-1} +\alpha \mu_q^2 \Sigma_q^{-1}x^T \Sigma_l^{-1}x - (1-\alpha)(x^T \Sigma_l^{-1}o)^2 -2\alpha \mu_q \Sigma_q^{-1}x^T \Sigma_l^{-1}o}{\alpha \Sigma_q^{-1} + (1-\alpha)( \Sigma_p^{-1} + x^T\Sigma_l^{-1}x)}   \right]=\\
& \frac{1}{2} \left[\frac{\alpha \Sigma_q^{-1}o^T\Sigma_l^{-1}o +(1-\alpha) \Sigma_p^{-1} o^T\Sigma_l^{-1}o + (1-\alpha)o^T\Sigma_l^{-1}o x^T\Sigma_l^{-1}x +\alpha \mu_q^2 \Sigma_q^{-1}\Sigma_p^{-1} +\alpha \mu_q^2 \Sigma_q^{-1}x^T \Sigma_l^{-1}x - (1-\alpha)(x^T \Sigma_l^{-1}o)^2 -2\alpha \mu_q \Sigma_q^{-1}x^T \Sigma_l^{-1}o}{\alpha \Sigma_q^{-1} + (1-\alpha)( \Sigma_p^{-1} + x^T\Sigma_l^{-1}x)}   \right]=\\
& \frac{1}{2} \left[\frac{\alpha \Sigma_q^{-1}o^T\Sigma_l^{-1}o +(1-\alpha) \Sigma_p^{-1} o^T\Sigma_l^{-1}o +\alpha \mu_q^2 \Sigma_q^{-1}\Sigma_p^{-1} +\alpha \mu_q^2 \Sigma_q^{-1}x^T \Sigma_l^{-1}x -2\alpha \mu_q \Sigma_q^{-1}x^T \Sigma_l^{-1}o}{\alpha \Sigma_q^{-1} + (1-\alpha)( \Sigma_p^{-1} + x^T\Sigma_l^{-1}x)}   \right]=\\
& \frac{\alpha}{2 \Sigma_q \Sigma_{\alpha}^{-1}} \left[o^T\Sigma_l^{-1}o +\Sigma_q \frac{(1-\alpha)}{\alpha} \Sigma_p^{-1} o^T\Sigma_l^{-1}o +\mu_q^2\Sigma_p^{-1} +\mu_q^2 x^T \Sigma_l^{-1}x -2\mu_q x^T \Sigma_l^{-1}o\right]\\
\end{align*}
\normalsize
The final for of the bound is:
\begin{align*}
D_{\alpha}(q(s)||p(s,o)) & = \frac{1}{2}\log\left(\frac{|\Sigma_q|}{(2\pi)^n |\Sigma_p| |\Sigma_l|}\right) - \\
    & \frac{\alpha}{2(\Sigma_q \Sigma_{\alpha}^{-1})}\left(o^T\Sigma_l^{-1}o + \mu_q^2 \Sigma_p^{-1} + \mu_q^2 x^T \Sigma_l^{-1} x -2\mu_q x^T \Sigma_l^{-1} o\right) -\\
    & \frac{1}{2(1-\alpha)} \log \left(1+(1-\alpha)(\Sigma_qx^T \Sigma_l^{-1} x + \Sigma_q\Sigma_p^{-1} - 1)    \right) - \\
    & \frac{1}{2 \Sigma_{\alpha}^{-1}}\left((1-\alpha)\Sigma_p^{-1} o^T \Sigma_l^{-1}o \right) =\\
\end{align*}
When alpha tends to 1, the first term is the same; the second is scaled; the third is a known limit; the last tends to zero. In general for different alpha the first 3 terms are scaled in a different way. Some are scaled with a log. The only new term is the last one, which does not depend on $\mu_q$
\section{Random things}
\begin{itemize}
    \item $\Sigma_q \Sigma_{\alpha}^{-1} = 1+(1-\alpha)(\Sigma_q\Sigma_{KL}^{-1})$
    \item If $\Sigma_q = \Sigma_{\alpha}$, then $\Sigma_q^{-1} = \Sigma_p^{-1} + x^T \Sigma_l^{-1}x$
    \item If $\Sigma_q = \Sigma_{KL}$, then $\Sigma_{\alpha}^{-1} = (\Sigma_p^{-1} + x^T \Sigma_l^{-1}x) (1- \frac{\alpha}{2})$ and $\Sigma_q \Sigma_{\alpha}^{-1} = 2 - \alpha$
\end{itemize}